\documentclass[fleqn,usenatbib]{mnras}
\usepackage[T1]{fontenc}

\usepackage[normalem]{ulem} 

\usepackage{graphicx}	
\usepackage{amsmath}	

\usepackage{color,soul}
\definecolor{lightblue}{rgb}{.70,.95,1}
\sethlcolor{lightblue} 

\usepackage{xspace} 
\renewcommand{\AA}{\normalfont\r{A}\xspace} 

\newcommand{\teff}{\ensuremath{T_{\mathrm{eff}}}\xspace}
\newcommand{\kms}{\ensuremath{\rm{km}\,s^{-1}}\xspace}
\newcommand{\logg}{\ensuremath{\log g}\xspace}
\newcommand{\feh}{\rm{[Fe/H]}\xspace}
\newcommand{\cfe}{\rm{[C/Fe]}\xspace}

\newcommand{\alphafe}{\rm{[\ensuremath{\alpha}/Fe]}\xspace}
\newcommand{\Gaia}{\textit{Gaia}\xspace}
\newcommand{\CaHK}{\emph{CaHK}\xspace}
\newcommand{\Pristine}{\emph{Pristine}\xspace}

\defcitealias{arentsen20b}{A20b}

\usepackage{newtxtext,newtxmath}

\title[PIGS III: CEMP stars in the bulge]{The Pristine Inner Galaxy Survey (PIGS) III: carbon-enhanced metal-poor stars in the bulge\thanks{based on observations made with the Canada-France-Hawaii Telescope (CFHT) and the Anglo-Australian Telescope (AAT)}}
 
\author[A. Arentsen et al.]{Anke Arentsen,$^{1}$\thanks{E-mail: anke.arentsen@astro.unistra.fr}
Else Starkenburg,$^{2}$
David S. Aguado,$^{3}$
Nicolas F. Martin,$^{1,4}$
Vinicius M. Placco,$^{5}$
\newauthor
Raymond Carlberg,$^{6}$,
Jonay I. Gonz\'alez Hern\'andez,$^{7,8}$
Vanessa Hill,$^{9}$
Pascale Jablonka,$^{10,11}$
\newauthor
Georges Kordopatis,$^{9}$
Carmela Lardo,$^{12}$
Lyudmila I. Mashonkina,$^{13}$
Julio F. Navarro,$^{14}$
Kim A. Venn,$^{14}$
\newauthor
Sven Buder$^{15,16}$, Geraint F. Lewis$^{17}$, Zhen Wan$^{17}$, Daniel B. Zucker$^{18,19}$
\\
\\
$^{1}$Universit\'e de Strasbourg, CNRS, Observatoire astronomique de Strasbourg, UMR 7550, F-67000 Strasbourg, France\\
$^{2}$Kapteyn Astronomical Institute, University of Groningen, Postbus 800, 9700 AV, Groningen, the Netherlands\\
$^{3}$Institute of Astronomy, University of Cambridge, Madingley Road, Cambridge CB3 0HA, UK \\
$^{4}$Max-Planck-Institut f\"ur Astronomie, K\"onigstuhl 17, D-69117 Heidelberg, Germany\\
$^{5}$Community Science and Data Center/NSF's NOIRLab, 950 N. Cherry Ave., Tucson, AZ 85719, USA \\
$^{6}$Department of Astronomy \& Astrophysics, University of Toronto, Toronto, ON M5S 3H4, Canada\\ 
$^{7}$Instituto de Astrof\'isica de Canarias, V\'ia L\'actea, 38205 La Laguna, Tenerife, Spain \\
$^{8}$Universidad de La Laguna, Departamento de Astrof\'isica, 38206 La Laguna, Tenerife, Spain \\
$^{9}$Universit\'e C\^ote d'Azur, Observatoire de la C\^ote d'Azur, CNRS, Laboratoire Lagrange, Nice, France \\
$^{10}$Institute of Physics, Laboratoire d'astrophysique, \'Ecole Polytechnique F\'ed\'erale de Lausanne (EPFL), Observatoire, CH-1290 Versoix, Switzerland \\
$^{11}$GEPI, Observatoire de Paris, Universit\'e PSL, CNRS, Place Jules Janssen, F-92190 Meudon, France \\ 
$^{12}$Dipartimento di Fisica e Astronomia, Università degli Studi di Bologna, Via Gobetti 93/2, I-40129 Bologna, Italy \\
$^{13}$Institute of Astronomy of the Russian Academy of Sciences, Pyatnitskaya st. 48, 119017, Moscow, Russia \\ 
$^{14}$Department of Physics \& Astronomy, University of Victoria, Victoria, BC, V8W 3P2, Canada \\
$^{15}$Research School of Astronomy \& Astrophysics, Australian National University, ACT 2611, Australia\\
$^{16}$Center of Excellence for Astrophysics in Three Dimensions (ASTRO-3D), Australia\\
$^{17}$Sydney Institute for Astronomy, School of Physics, A28, The University of Sydney, NSW 2006, Australia \\
$^{18}$Department of Physics and Astronomy, Macquarie University, Sydney, NSW 2109, Australia \\
$^{19}$Macquarie University Research Centre for Astronomy, Astrophysics \& Astrophotonics, Sydney, NSW 2109, Australia
}

\date{Accepted 2021 May 7. Received 2021 May 6; in original form 2021 April 14}

\pubyear{2021}

\begin{document}
\label{firstpage}
\pagerange{\pageref{firstpage}--\pageref{lastpage}}
\maketitle

\begin{abstract}
The most metal-deficient stars hold important clues about the early build-up and chemical evolution of the Milky Way, and carbon-enhanced metal-poor (CEMP) stars are of special interest. However, little is known about CEMP stars in the Galactic bulge. 
In this paper, we use the large spectroscopic sample of metal-poor stars from the \Pristine Inner Galaxy Survey (PIGS) to identify CEMP stars ($\cfe \geqslant +0.7$) in the bulge region and to derive a CEMP fraction. 
We identify 96 new CEMP stars in the inner Galaxy, of which 62 are very metal-poor ($\feh < -2.0$); this is more than a ten-fold increase compared to the seven previously known bulge CEMP stars. 
The cumulative fraction of CEMP stars in PIGS is $42^{\,+14\,}_{\,-13} \%$ for stars with $\feh < -3.0$, and decreases to $16^{\,+3\,}_{\,-3} \%$ for $\feh < -2.5$ and $5.7^{\,+0.6\,}_{\,-0.5} \%$ for $\feh < -2.0$. 
The PIGS inner Galaxy CEMP fraction for $\feh < -3.0$ is consistent with the halo fraction found in the literature, but at higher metallicities the PIGS fraction is substantially lower. 
While this can partly be attributed to a photometric selection bias, such bias is unlikely to fully explain the low CEMP fraction at higher metallicities.  
Considering the typical carbon excesses and metallicity ranges for halo CEMP-s and CEMP-no stars, our results point to a possible deficiency of both CEMP-s and CEMP-no stars (especially the more metal-rich) in the inner Galaxy. The former is potentially related to a difference in the binary fraction, whereas the latter may be the result of a fast chemical enrichment in the early building blocks of the inner Galaxy. 
\end{abstract}



\textcolor{white}{.}
\vspace{4cm}

\section{Introduction}

The Milky Way contains stars with a wide range of metallicities, covering different formation time scales and environments. The most metal-poor stars are typically expected to be among the oldest, and are mainly found in low-density environments such as the Galactic halo and dwarf galaxies. Metal-poor stars contain important clues about the First Stars and the earliest formation history of the Milky Way, which is why they have been searched for and studied extensively \citep[see, e.g.,][for a recent review]{frebelnorris15}. 

A region in the Milky Way largely missing from the study of metal-poor stars has been the Galactic bulge. The halo is expected to have its highest density in the inner Galaxy, but it is an extremely challenging area to study. Metal-poor stars are highly outnumbered by metal-rich stars, and dust extinction obstructs and disturbs our view. However, thanks to efficient pre-selection methods and/or large surveys, the number of very metal-poor inner Galaxy stars identified and studied has been increasing in recent years \citep{ness13a, garciaperez13, howes14, howes15, howes16, caseyschlaufman15, koch16, lamb17, lucey19, lucey21, reggiani20, arentsen20b}. 

One striking property of the most iron-poor stars in the halo is that they often show very large over-abundances of carbon, with the fraction of such carbon-enhanced metal-poor (CEMP) stars increasing with decreasing metallicity \citep[e.g.][]{beerschristlieb05,yong13}. In the Galactic halo, these CEMP stars are found to comprise 30\% of stars with \feh\footnote{[X/Y] $ = \log(N_\mathrm{X}/N_\mathrm{Y})_* - \log(N_\mathrm{X}/N_\mathrm{Y})_{\odot}$, where the asterisk subscript refers to the considered star, and N is the number density. In this work, \feh and metallicity are used interchangably.} $< -2$ and up to 80\% of stars with $\feh < -4$ \citep{placco14, yoon18}, assuming a CEMP definition of $\cfe \geqslant +0.7$. 

Many of the CEMP stars with higher metallicities ($\feh > -3.0$) show enhancements in s-process elements and a very high absolute carbon abundance, fitting with the hypothesis that they are chemically enriched by mass transfer from a companion star that has gone through the asymptotic giant branch (AGB) phase \citep[e.g.,][]{lucatello05,bisterzo10,abate15a}. Because of their high s-process signature, these stars are referred to as CEMP-s stars \citep{beerschristlieb05}. Indeed, a very high fraction of these stars have radial velocity variations consistent with them being in binary systems \citep[][]{lucatello05,hansen16b}. 

Most CEMP stars, however, show carbon enhancement without the accompanying s-process signature \citep[for an overview see e.g.][]{norris13b}. They are dominant among the most iron-poor stars, have lower absolute carbon values \citep{spite13,yoon16} and they are typically not in binary systems (\citealt{starkenburg14,hansen16a}, although there are some interesting exceptions described in those works and in \citealt{arentsen19_cemp}). These CEMP-no stars were likely born with their carbon over-abundance, out of the early interstellar medium enriched by the First Stars. Two possible hypotheses for why the First Stars produce so much carbon are that they might have higher spin rates  \citep{chiappini06, meynet06} and/or that they might explode as mixing and fallback (faint) supernovae rather than ordinary supernovae \citep{umedanomoto03, nomoto13, tominaga14}. 

The fraction of CEMP stars and the relative number of CEMP-s versus CEMP-no stars appear to vary throughout the Milky Way and its satellite galaxies. Relatively small samples of several hundred very metal-poor stars studied in detail have indicated trends of increasing CEMP fraction with increasing height from the Galactic plane \citep[][]{frebel06}, and a larger relative number of CEMP-no stars further away into the halo \citep[][]{carollo14}. Similar results are seen in recent work using tens of thousands of stars with low resolution spectroscopy from the Sloan Digital Sky Survey (SDSS) and the AAOmega Evolution of Galactic Structure (AEGIS) survey, which produced ``carbonicity'' maps and found the average \cfe to increase with increasing distance from the Sun \citep{lee17, lee19, yoon18}. They also found that the relative number of CEMP-s stars is larger closer to the Sun, and that CEMP-no stars appear to become more dominant at larger distances. In the Milky Way satellites, an apparent lack of CEMP-no stars has been observed in dwarf spheroidal galaxies, whereas the CEMP-no stars are found in ``usual'' frequencies in ultra faint dwarf galaxies \citep[e.g.][]{norris10, lai11, frebel14, skuladottir15}. Because of these observations, some of these studies have suggested that the fraction of CEMP-no stars may be lower in more massive building blocks. However, others have suggested that this is simply a selection effect related to the varying metallicity distribution functions in the different types of dwarf galaxies \citep{salvadori15}. The carbonicity gradient in the halo can thus be related to the halo metallicity gradient and/or the type of building block that contributed to each part. To summarise, spatial metallicity and carbon trends can give us valuable insight into the early chemical and dynamical evolution of the Milky Way and its building blocks. 

The Extremely Metal-poor BuLge stars with AAOmega (EMBLA) survey \citep{howes14,howes15,howes16} was the first large metal-poor bulge survey, using metallicity-sensitive SkyMapper photometry \citep{bessell11, wolf18} to select the most metal-poor candidates for spectroscopic follow-up. They derived the carbon abundances for 33 metal-poor stars (almost all $\feh < -2.0$), and found only one of them to have a predicted natal \cfe (corrected for stellar evolution effects) above $\cfe = +1.0$. This is in striking contrast with the much higher fraction of CEMP stars in the Galactic halo. A concern with that result is that the photometric SkyMapper selection could be biased against carbon-rich stars \citep[shown in e.g.][]{dacosta19,chiti20}, since the wavelength region covered by their metallicity-sensitive narrow-band $v$ filter also contains a CN band. Given its major implications for early chemical evolution in our Galaxy, it is of importance that the inner Galaxy CEMP fraction is further studied with larger sample sizes and independent methods. Other previous metal-poor bulge studies either did not measure carbon and/or had very small sample sizes. 

In this work, we present a complementary view of carbon abundances in the inner Galaxy from the \Pristine Inner Galaxy Survey \citep[PIGS,][hereafter A20b]{arentsen20b}. Similar to EMBLA, PIGS also makes use of metallicity-sensitive photometry to select candidate stars, but a difference is that the employed narrow-band filter is significantly narrower than the SkyMapper $v$ filter -- it is therefore expected to be less affected by surrounding CN and CH molecular bands. We make use of the low/intermediate-resolution spectroscopic follow-up of thousands of PIGS stars with measured metallicities and carbon abundances to study CEMP stars in the inner Galaxy. The relevant details about PIGS are presented in Section~\ref{sec:pigs}. We then describe the behaviour of carbon on the red giant branch in the PIGS sample and derive evolutionary carbon corrections for the PIGS \cfe determinations in Section~\ref{sec:carbongiants}. In Section~\ref{sec:cemp} we present the discovery of many new CEMP stars in the inner Galaxy, and study the CEMP fraction as function of \feh. We discuss the interpretations and limitations of our results in Section~\ref{sec:disc}, and summarise our conclusions in Section~\ref{sec:conc}.

\section{The \Pristine Inner Galaxy Survey (PIGS)}\label{sec:pigs}

\subsection{Survey overview}

\begin{figure}
\centering
\includegraphics[width=0.93\hsize,trim={0.0cm 0.0cm 0.0cm 0.0cm}]{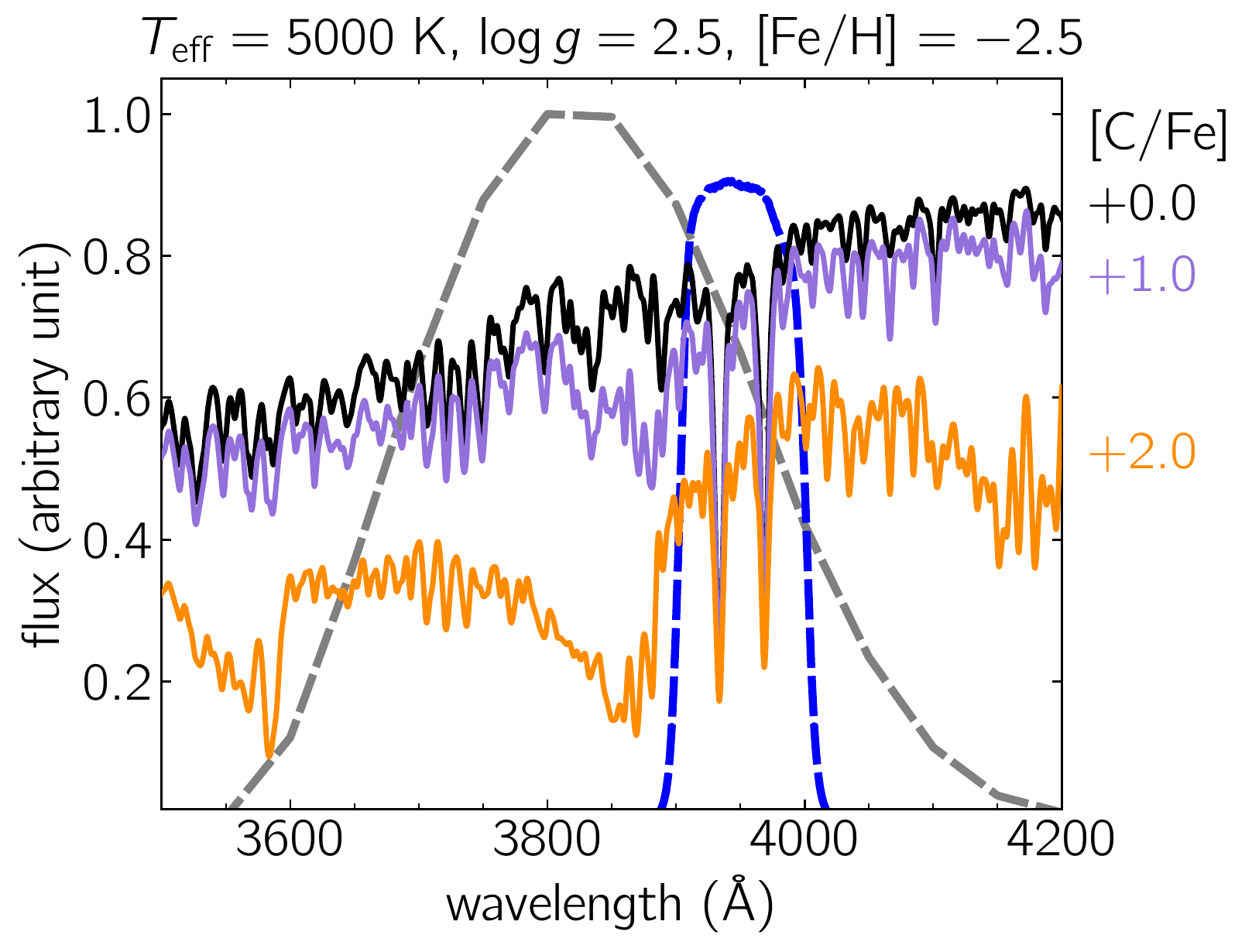} 
\caption{Filter curves for the SkyMapper $v$ (grey) and \Pristine \CaHK (blue) filters, plotted over synthetic spectra for three stars with the same stellar parameters (listed above the figure), but different [C/Fe]. The CNO abundances are fixed to ``typical'' values for CEMP-no stars (see the text for details). The \Pristine filter is much narrower than the SkyMapper filter, and is expected to be less influenced by enhanced carbon.}
    \label{fig:carbonspec}
\end{figure}

For details on the \Pristine Inner Galaxy Survey we refer the reader to \citetalias{arentsen20b}, the second paper in the PIGS series describing the survey photometry, selection for follow-up and spectroscopic analysis. Here, we only briefly summarise the relevant details. PIGS is an extension of the main \Pristine survey, a photometric survey using  the metallicity-sensitive narrow-band \CaHK MegaCam filter on the Canada-France-Hawaii-Telescope (CFHT) to find and study the most metal-poor stars \citep[for an overview, see][]{starkenburg17b}. The \CaHK filter centred on the strong Ca H\&K stellar absorption lines is shown in Figure~\ref{fig:carbonspec} together with the broader SkyMapper $v$-filter, on top of synthetic spectra of varying carbon abundance. The synthetic spectra were created using publicly available stellar atmosphere models from MARCS \citep{Gustafsson08,Plez08}, combined with the Turbospectrum spectral synthesis code \citep{alvarez98}. We assumed [C/N] = 0, and [O/Fe] increasing with increasing \cfe (a typical signature for CEMP-no stars): [O/Fe] = +0.6 for carbon-normal stars and $+1.3$ and $+1.9$ for $\cfe = +1.0$ and $+2.0$, respectively. 

The \Pristine survey has an extensive spectroscopic follow-up campaign of its most promising candidates \citep[][]{youakim17,caffau17,starkenburg18,aguado19,bonifacio19,venn20,caffau20,kielty20}, and it has a dedicated dwarf galaxy program  \citep{longeard18,longeard20b,longeard20a}. Instead of focusing on the Galactic halo and the satellites that reside in it, PIGS is looking towards the inner regions of the Galaxy -- a far more challenging environment for a number of reasons. First, the bulge is on average much more metal-rich, having a metallicity distribution that peaks only slightly below solar \citep[around $\sim-0.2/-0.5$ depending on the latitude, e.g.][]{ness13a}, whereas in the halo the mean metallicity lies around $\feh = -1.6$ \citep[e.g.,][]{youakim20}. Secondly, the amount of dust and therefore of extinction is far greater towards the inner Galaxy than towards the Galactic halo, and it can be very inhomogeneous. The limited resolutions of available extinction maps and variations of the extinction law across the sky \citep{schlafly16} severely hamper an accurate extinction correction. The bulge footprint of PIGS is chosen such that it avoids the most extincted regions; it surveys regions with E(B$-$V)~$\lesssim 0.7$ for the part of the sky bound by declination $> -30^{\circ}$ (because CFHT is a Northern Hemisphere facility) and $l$ and $b$ roughly between $-12^{\circ}$ and $+12^{\circ}$.

\begin{figure}
\centering
\includegraphics[width=1.0\hsize,trim={0.0cm 0.0cm 0.0cm 0.0cm}]{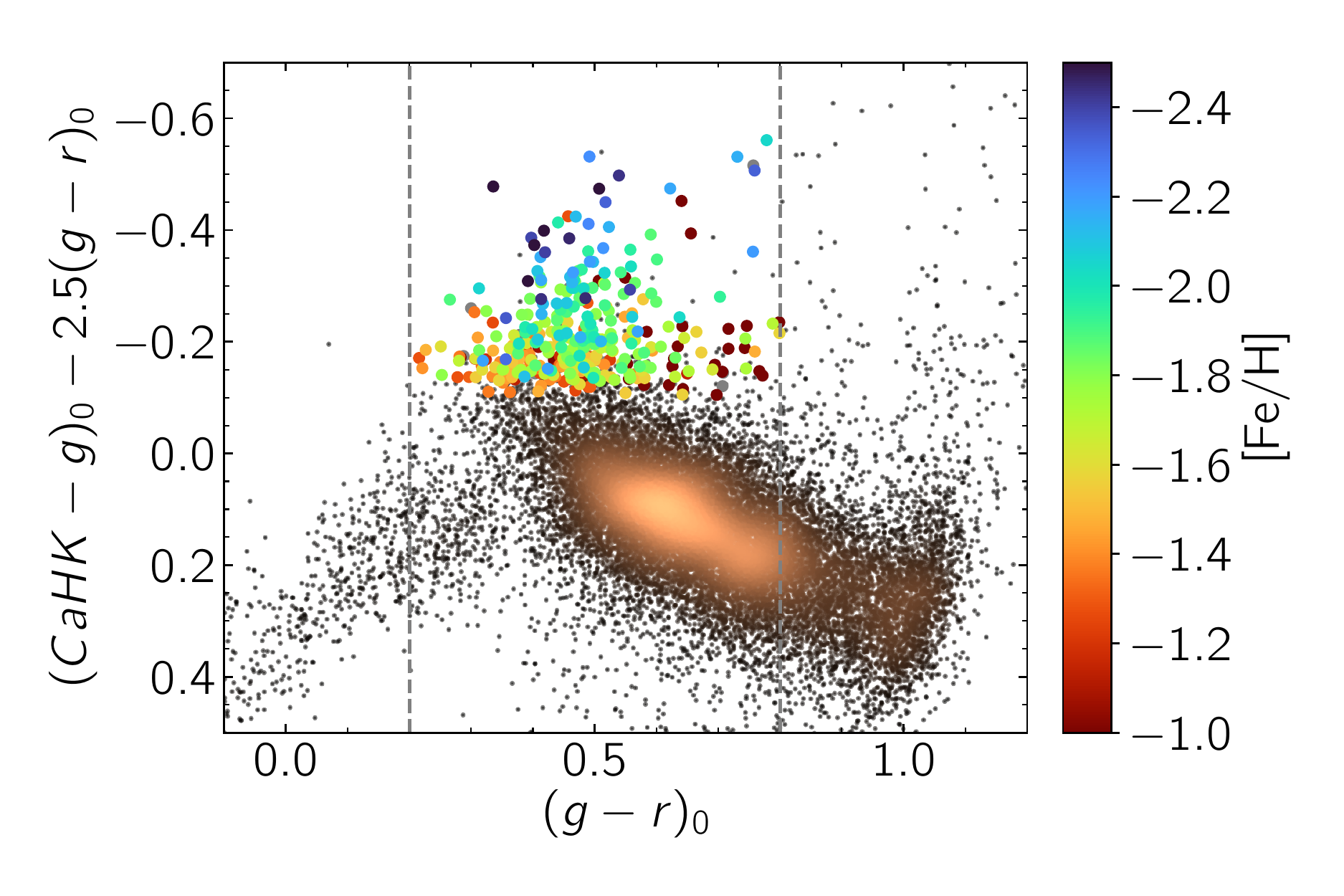}
\caption{PIGS colour-colour diagram for a follow-up field selected with PS1. The field is located at $(l,b) = (7.6^{\circ}, 8.0^{\circ})$, has a radius of 1$^{\circ}$ and has an average E(B$-$V) = 0.51. The background shows the density of all $\sim 22\,000$ stars passing our quality selection criteria in this field; the $\sim$350 metal-poor candidates followed-up with AAT have been colour-coded by their spectroscopic \feh. The vertical lines indicate the colour cuts made for the follow-up selection.}
    \label{fig:ccd}
\end{figure}

Much of the PIGS footprint has been followed-up with spectroscopic observations using AAOmega+2dF at the AAT  \citep{saunders04,lewis02, sharp06}. As discussed in \citetalias{arentsen20b}, the exact details of the target selection have varied throughout the progression of the survey. In all instances, however, the targets are selected from a colour-colour space including the \CaHK and broad-band colours either from \Gaia DR2 \citep{gaia18} or from PanSTARRS1 \citep[PS1,][]{panstarrs}. The colour-colour selection for an example field is shown in Figure~\ref{fig:ccd}. The broad-band photometry is necessary as a proxy for the temperature of the star, and a combination of broad-band colours and \Pristine \CaHK photometry is used as an indication of the metallicity. Because of unavoidable difficulties with differential reddening and the zero-point of the \CaHK photometry in this challenging Galactic region, we refrained from calculating photometric metallicities for each star. Instead, candidates were selected on a relative basis in each field and the fibres of the AAOmega spectrograph were filled from lower to higher expected metallicities, starting from the most metal-poor candidates located in the upper part of the colour-colour diagram. \citetalias{arentsen20b} already showed that this selection has an unprecedentedly high efficiency in selecting the most metal-poor populations in the inner Galaxy: 90\% of the best candidates are confirmed to satisfy \feh $< -2.0$ from the spectroscopy, as do 60-75\% of the next best selection box (depending on the magnitude of extinction in the field). 

\begin{figure}
\centering
\includegraphics[width=1.0\hsize,trim={0.0cm 0.0cm 0.0cm 0.0cm}]{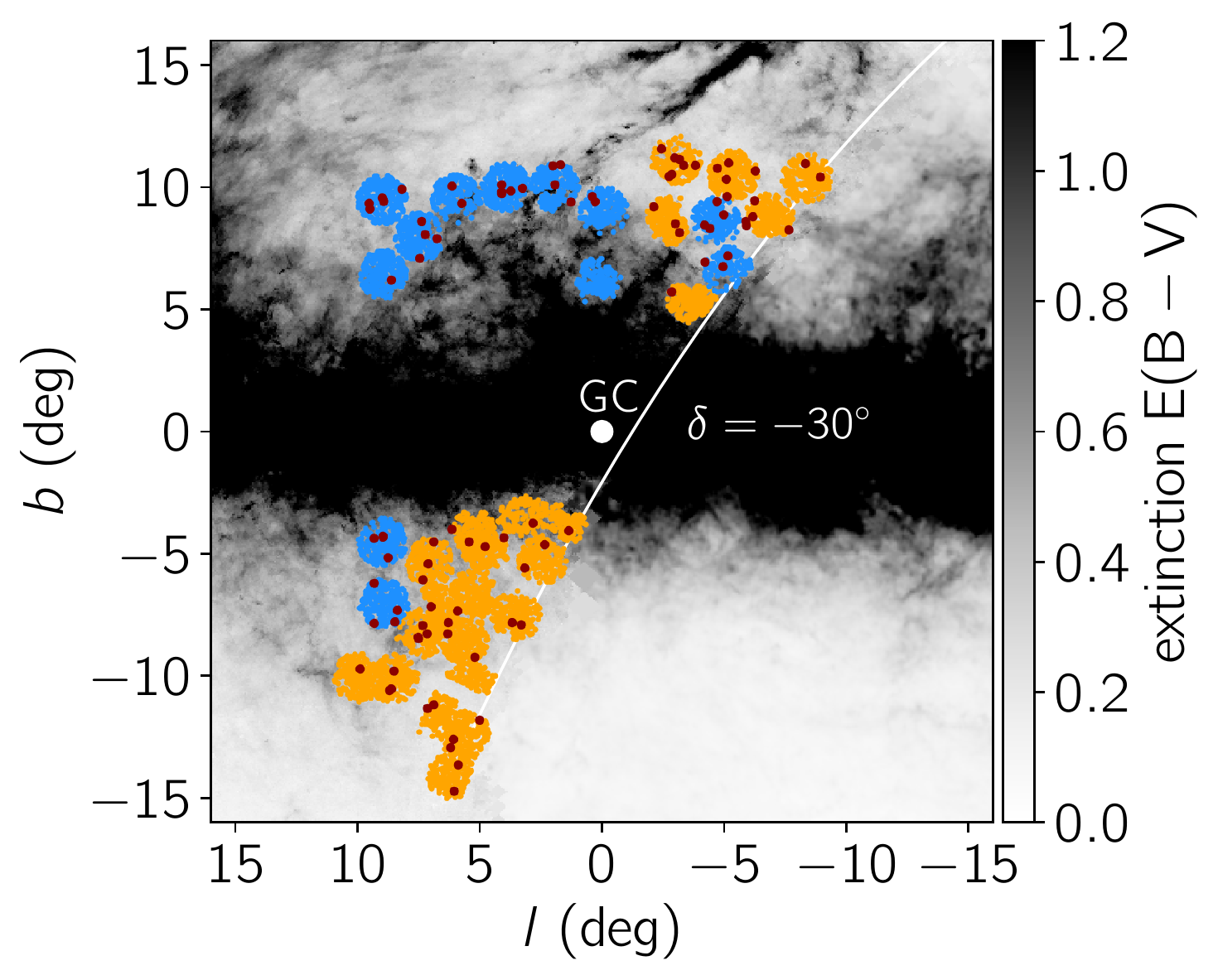} 
\caption{PIGS spectroscopic footprint for stars with $\feh < -1.5$, with orange fields already present in \citetalias{arentsen20b} and blue fields added in 2020. CEMP stars from this work are indicated as dark red points. The extinction comes from \citet{green19} above $\delta = -30^{\circ}$ and from \citet{schlegel98} below. }
    \label{fig:foot}
\end{figure}

The first paper of the PIGS series \citep{arentsen20a} used the spectroscopic follow-up observations to investigate the kinematical signature of metal-poor ($\feh < -1.0$) inner Galaxy stars and compared it with the stars at higher metallicities. It shows for the first time how the signature of Galactic rotation, observed in metal-rich stars, decreases with metallicity for metal-poor stars and completely disappears for stars with $\feh < -2.0$.

As we selected stars with a limited magnitude range of $13.5 < G < 16.5$ for \Gaia, or $14.0 < g < 17.0$ for PS1, and we deselect foreground dwarf stars using \Gaia parallax information (see \citetalias{arentsen20b}), the stars are expected to be roughly at the distance of the bulge. A kinematic analysis with \Gaia data investigating the detailed orbits of the PIGS stars is planned for a future work. 

The most recent PIGS follow-up coverage is shown in Figure~\ref{fig:foot}, which includes 12 additional fields (observed in 2020) compared to \citetalias{arentsen20b}. Altogether, the PIGS AAOmega+2dF spectroscopic follow-up sample contains $\sim$12\,000 stars with low/medium-resolution spectra covering $3700-5500$~\AA\ (blue arm, R$\sim$1300) and $8400-8800$~\AA (red arm, R$\sim$11\,000). The final sample (after quality cuts) contains $\sim4800$ intermediate metal-poor (IMP) stars with $-2.0 < \feh < -1.5$, and $\sim1900$ very metal-poor (VMP) stars with $\feh < -2.0$. The median signal-to-noise ratio (SNR) per pixel is 19, 57 and 50 for $4000-4100$~\AA, $5000-5100$~\AA~and $8400-8800$~\AA, respectively. 

\subsection{Stellar parameters and carbon abundances}\label{sec:ferre}

The PIGS spectra have been analysed with two independent pipelines (FERRE\footnote{FERRE \citep{allende06} is available from \url{http://github.com/callendeprieto/ferre}} and ULySS\footnote{ULySS \citep{koleva09} is available from \url{http://ulyss.univ-lyon1.fr/}}), which yielded largely consistent results \citepalias{arentsen20b}. In this work we will use the FERRE analysis, since it includes \cfe and extends to lower metallicities than the ULySS analysis. The method will be briefly summarised here.

\begin{figure}
\centering
\includegraphics[width=1.0\hsize,trim={0.0cm 0.0cm 0.0cm 0.0cm}]{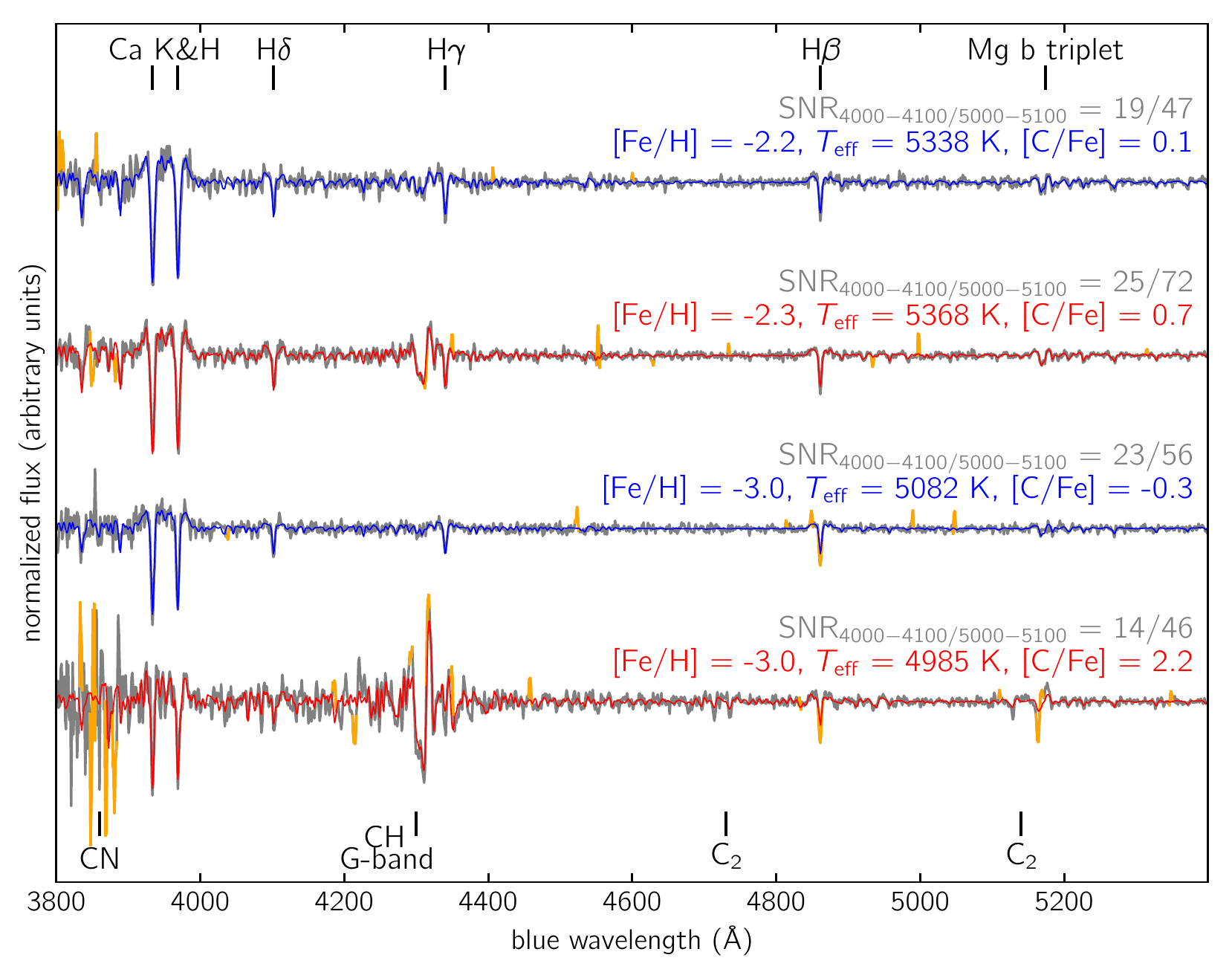}
\caption{Example FERRE fits for two carbon-normal stars (blue) and two CEMP stars (red). The top two spectra have roughly the same metallicity and temperature, and the bottom two spectra as well (but a different metallicity and temperature as the top two). A running mean normalisation has been applied to both the observed and model spectra, and orange regions indicate outlier pixels excluded from the fit.}
    \label{fig:specs}
\end{figure}

The full-spectrum fitting code FERRE was employed for the derivation of the effective temperature \teff, surface gravity \logg, metallicity \feh and the carbon abundance \cfe. We fit the observed spectra against a synthetic stellar model library, interpolating between the nodes with a cubic algorithm. The blue and red arm of the spectra are fit simultaneously to derive \teff, \logg, \feh and \cfe. The reference synthetic grid is an extension of the carbon-rich grid described in \citet{aguado17}, based on Kurucz model atmospheres \citep{meszaros12} computed with the ASSET code \citep{koesterke08}. The new grid includes cooler stars (down to \teff = 4500 K) and extends to higher metallicities (up to \feh = +0.5), with steps in \teff, \logg, \feh, and \cfe of 250~K, 0.5~dex, 0.5~dex, and 1.0~dex, respectively. The alpha abundances are assumed to be enhanced following $\alphafe = +0.4$, which is typically appropriate for the metal-poor populations we focus on in this work. The spectra were self-consistently calculated using carbon and alpha-enhanced atmospheres, adopting a microturbulence of $2\kms$ (appropriate for giants). Four example fits are shown in Figure~\ref{fig:specs}.

The comparison between the two analyses in \citetalias{arentsen20b} was used to derive final uncertainties on the stellar parameters. However, the ULySS analysis does not include \cfe as a parameter and, as such, we do not have a straightforward measurement of the external uncertainty to derive final uncertainties for \cfe. We find that the \emph{internal} uncertainties on \cfe are roughly two times larger than the internal uncertainties derived for \feh. Assuming that the external uncertainties on \cfe will follow the same trend, we quadratically add an uncertainty floor of $2 \times 0.11$ (the external \feh uncertainty) $= 0.22$~dex to all FERRE \cfe uncertainties. The median of the final \cfe uncertainties for our sample is 0.27~dex, and for \teff, \logg and \feh they are 148~K, 0.41~dex and 0.14~dex, respectively.

\begin{figure}
\centering
\includegraphics[width=0.8\hsize,trim={0.0cm 0.0cm 0.0cm 0.0cm}]{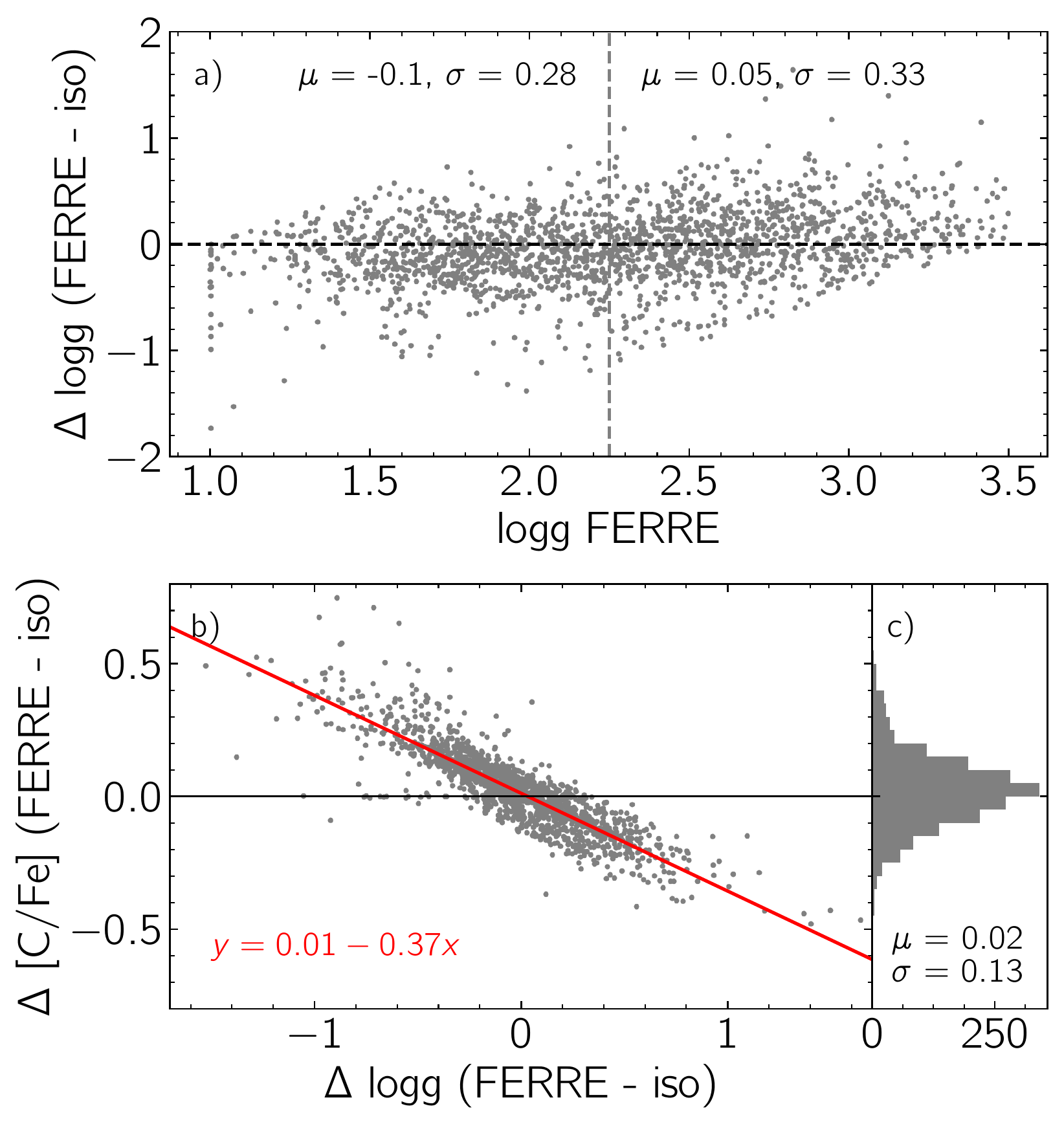} 
\caption{Tests of \logg and \cfe for VMP stars ($\feh < -2.0$) in PIGS. Panel a) presents the difference between FERRE gravities and isochrone gravities, as function of the FERRE gravity. Panel b) shows the correlation between a difference in \logg and the resulting difference in \cfe. A linear fit to the points is presented in red. Panel c) shows the distribution of the \cfe differences, with the mean and standard deviation indicated.}
    \label{fig:dloggcfe}
\end{figure}

\subsection{Degeneracy between carbon abundance and surface gravity}\label{sec:loggbias}

\citet{aguado19} noted that there is a degeneracy between \logg and \cfe in the parameter determination for VMP stars with FERRE. A bias in \logg could therefore result in a bias in \cfe. Their Figure~5 shows the difference in \cfe when they let \logg be a free parameter versus when they set \logg to a fixed value. There is an almost linear relation between the difference in \cfe and the \logg for $1.0 < \logg < 3.5$, with an increase of 0.2~dex in \cfe for a decrease in \logg of 0.5~dex. It is challenging to derive \logg from low/medium-resolution spectroscopy, especially for VMP stars since there are not many (gravity-sensitive) features in the spectrum. The PIGS \logg values are better determined than those from \citet{aguado19}, because the spectra have higher SNR on average and because of the addition of the calcium triplet (this improves the results especially for lower gravity stars, see \citetalias{arentsen20b}). However, it is still possible that there could be some bias in \logg, and therefore in \cfe. 

To test for possible surface gravity/carbon abundance biases in the PIGS VMP sample, we compare the spectroscopic \logg values to \logg values from isochrones, and we perform a second FERRE analysis where we adopt the isochrone \logg. We used Yonsei-Yale isochrones \citep{demarque04} of 12~Gyr in a grid from $\feh=-2.0$ to $-3.5$ with steps of 0.1~dex, and adopted the \logg of the point on the isochrone red giant branch (RGB) closest to the FERRE \feh and \teff. A minority of the PIGS stars are likely on the early AGB instead of the RGB, for which this procedure is not entirely valid. These stars will be discussed in more detail in Section~\ref{sec:carbongiants}. The difference between the \logg values from FERRE and from the isochrones is shown in panel a) of Figure~\ref{fig:dloggcfe}. They tend to agree well, with an average difference of $-0.1$~dex and $0.05~dex$ for the lower and upper RGB, respectively, and a standard deviation of $\sim 0.3$~dex around the average in both regimes. 

We then rerun FERRE for the VMP stars with only \cfe as a free parameter, fixing \logg to the isochrone value and fixing \teff and \feh to their values from the initial FERRE analysis. These two parameters are less sensitive to \logg. Panel b) of Figure~\ref{fig:dloggcfe} shows the correlation between the difference in \logg and the difference in the derived \cfe, and panel c) presents the distribution of the difference in \cfe. The correlation is similar as what was found by \citet{aguado19}, but the average $\Delta \cfe$ is only 0.02~dex (with a standard deviation of 0.13~dex). We therefore conclude that carbon abundances derived in this work are not biased due to biased surface gravities, and use only the purely spectroscopic parameters.

\section{Stellar evolution on the giant branch}\label{sec:carbongiants}

As stars evolve on the RGB, their surface carbon abundance changes. Since PIGS contains almost exclusively giant stars, we discuss this effect here. In the inner layers of an RGB star, carbon has been converted to nitrogen in the CNO cycle. While ascending the RGB, the star experiences mixing episodes, in which material from layers deeper in the star are transported to the stellar surface. The first dredge up occurs at the start of the RGB, which does not significantly change the surface \cfe for metal-poor stars \citep[][]{vandenberg88, charbonnel94, gratton00}. No additional mixing is expected to take place from standard theoretical models, but observations have shown that the surface abundance of carbon and nitrogen changes significantly on the RGB for metal-poor stars. The strength of this ``extra mixing'' appears to depend on the initial metallicity and carbon and nitrogen abundances, and it has been found to be stronger in metal-poor stars \citep[e.g.][]{gratton00, stancliffe09, placco14}. One proposed mechanism is thermohaline mixing, which has been shown to reproduce the abundances in low-mass, low-metallicity evolved giants relatively well \citep{stancliffe07, charbonnel10, denissenkov10}. Recently, \citet{shetrone19} used a large sample of stars from APOGEE to study metallicity-dependent mixing and extra mixing in field giant stars. They confirmed that the extra mixing is extremely sensitive to metallicity, and that current standard models cannot adequately explain this effect. In this section, we investigate the change of carbon abundance on the RGB in PIGS, and determine evolutionary \cfe corrections. 

\subsection{[C/Fe] in the Kiel diagram of PIGS}

\begin{figure}
\centering
\includegraphics[width=1.0\hsize,trim={0.0cm 0.0cm 0.0cm 0.0cm}]{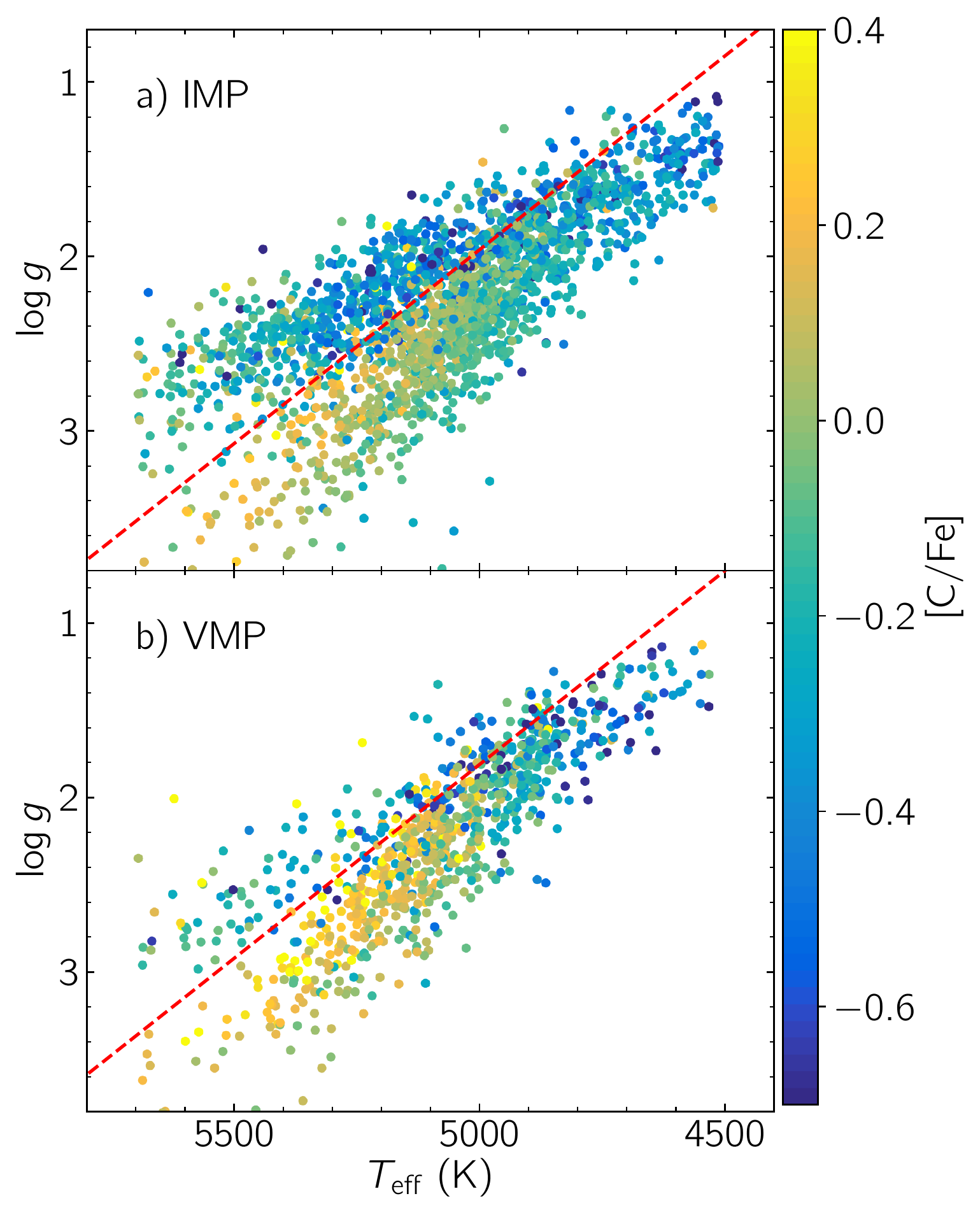} 
\caption{Kiel diagram of PIGS giants with SNR$_{4000-4100} > 25$, $\teff < 5700$~K and $\cfe < +0.7$, for stars with $-2.0 < \feh < -1.5$ (IMP, top panel) and $\feh < -2.0$ (VMP, bottom panel), colour-coded by \cfe. The dotted red lines indicate a rough separation between the RGB and AGB stars. }
    \label{fig:HR-carbon}
\end{figure}

It is possible to visualise the effect of extra mixing in PIGS on the Kiel diagram (\teff\,-- \logg). To this end, we select stars with SNR$_{4000-4100} > 25$ to only include those with good quality carbon abundances. The median total uncertainty on \cfe ($\epsilon_{\cfe}$) is 0.22~dex in this high SNR sample. The sample is split in two: the IMP stars with $-2.0 < \feh < -1.5$ and the VMP stars with $\feh < -2.0$. We present Kiel diagrams colour-coded by \cfe for these two samples in Figure~\ref{fig:HR-carbon}. The average value of \cfe clearly decreases for stars higher up the RGB (with lower \logg values), for both the IMP and the VMP stars. This is expected with what is known from the literature \citep[e.g.][]{gratton00, placco14}, although it has not been observed directly in the bulge region before. A comparison of depletion of carbon along the RGB in PIGS with literature models is discussed Section~\ref{sec:cfelogg}. 

There also appears to be a population of stars above the RGB that shows lower \cfe values (especially in the IMP sample). In the independent ULySS analysis of the spectra there are also two sequences in the \teff--\logg diagram. We hypothesise that these stars are in fact not RGB stars, but that they lie on the parallel early AGB instead. Early AGB stars are expected to have a lower carbon abundance than RGB stars because they have experienced extra mixing all the way up the RGB, before they became horizontal branch stars and subsequently AGB stars. We make a rough separation of both populations indicated by a solid line in Figure~\ref{fig:HR-carbon} (slightly different for the IMP and VMP stars). The relative number of AGB stars is discussed in Section~\ref{sec:agb}.  

\subsubsection{Carbon depletion with \logg}\label{sec:cfelogg}

\begin{figure}
\centering
\includegraphics[width=1.0\hsize,trim={0.0cm 0.8cm 0.0cm 0.0cm}]{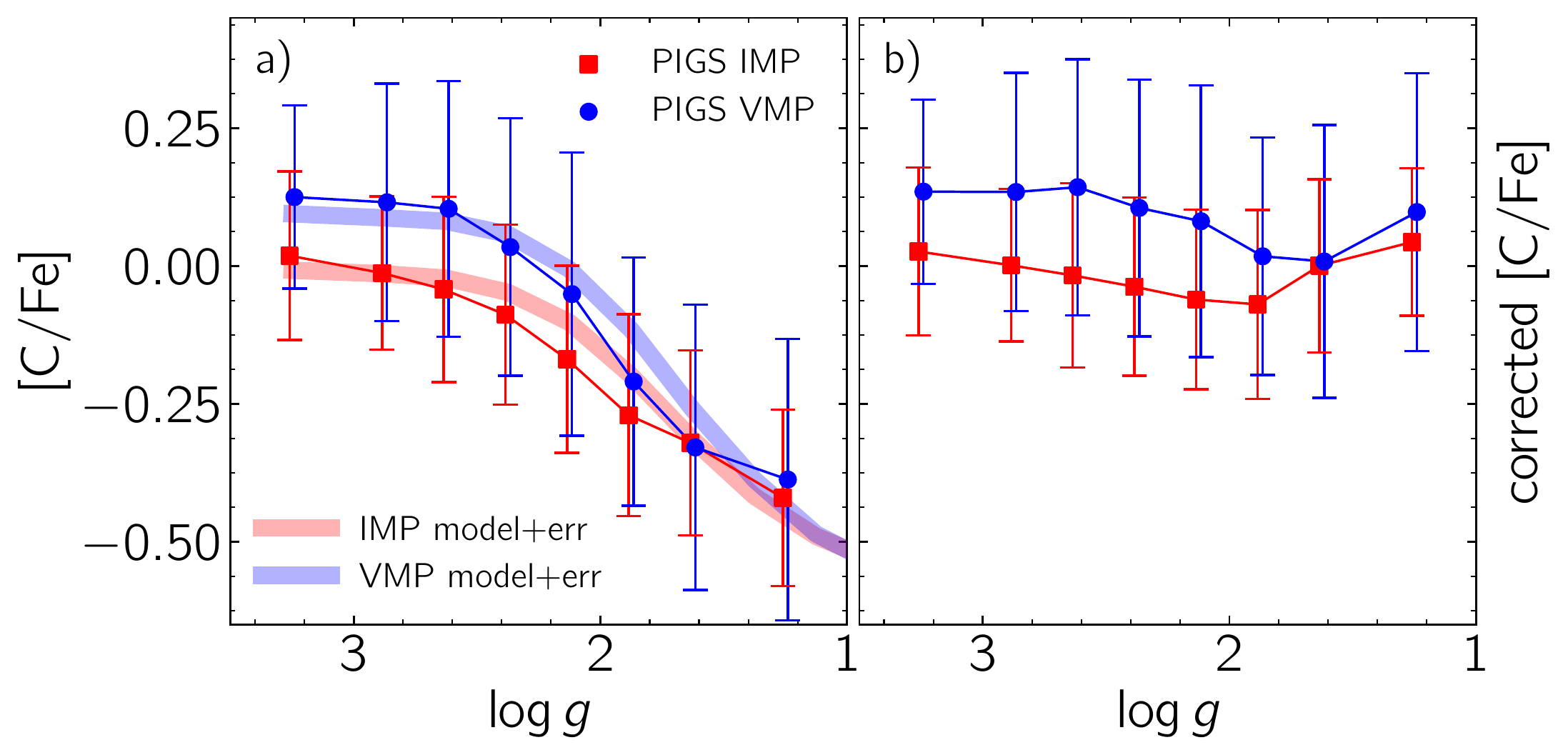} 
\caption{Left: mean \cfe versus \logg in the IMP and VMP samples of RGB stars (stars below the dotted lines in Figure~\ref{fig:HR-carbon}). The means are determined using a $5\sigma$ clipping of outliers, the error bars represent the standard deviation in each \logg bin. Also shown are theoretical predictions from the models by \citet{placco14}, assuming typical parameters and uncertainties for stars in our IMP and VMP samples (see the main text for details). Right: mean \cfe corrected for evolutionary effects versus \logg. }
    \label{fig:loggC}
\end{figure}

Panel a) of Figure~\ref{fig:loggC} presents the mean trends of \cfe with \logg for the IMP and VMP RGB stars. The trend of decreasing \cfe with the ascension on the RGB is clearly visible. Additionally, we find that the average \cfe for the unevolved VMP stars is $\sim 0.1$~dex higher than for the IMP stars. A trend of increasing \cfe with decreasing \feh is also present in metal-poor halo samples (e.g., \citealt{roederer14, amarsi19}), although the latter authors show that the trend could largely be the result of 3D/non-local thermodynamic equilibrium (non-LTE) effects. Most literature studies, however, do not correct for 3D and/or non-LTE effects. We therefore assume that it is safe to compare our results with those of others, as long as the stars are in a similar range of the parameter space. 

Also shown are predictions for the extra depletion of carbon in RGB stars, from the models by \citet{placco14} smoothed by the PIGS uncertainties. For the IMP stars we adopted $\feh = -1.75$ and natal $\cfe = 0.0$, and for the VMP stars $\feh = -2.3$ and natal $\cfe = +0.1$. In the original \citet{placco14} models, the extra depletion of carbon starts around $\logg = 2.0$ and increases steeply with decreasing \logg. However, the uncertainties of \cfe and especially \logg in PIGS are significant, spreading out this sharp feature. We took one thousand draws around each of the original model points from normal distributions with standard deviations equal to the typical PIGS uncertainties in \logg and \cfe to mimic this effect. For the high SNR sample used here, those uncertainties are 0.37~dex and 0.22~dex for \logg and \cfe, respectively. After this correction for the uncertainties, the models and the data agree well with each other. The  carbon depletion appears to be slightly higher in the data than in the models for $2.5 > \logg > 1.7$, which could be the result of some (carbon-poor) AGB contamination on the RGB in this \logg range (see also Figure~\ref{fig:HR-carbon}). 

We confirm that the evolutionary carbon depletion in our observations follows the theoretical predictions from \citet{placco14}. The derivation of evolutionary carbon corrections for each individual star in the PIGS sample will be described in Section~\ref{sec:ccor}. 

\subsubsection{Asymptotic giant branch (AGB) stars}\label{sec:agb}

In the IMP sample, the total number of RGB stars is roughly the same as the number of AGB stars. Given that stars generally live longer on the RGB than on the AGB, this seems like a large fraction of AGB stars at face value. To gain a more quantitative understanding, we roughly estimate the expected occurrence ratio of such stars using two models: a MIST evolutionary track\footnote{\url{http://waps.cfa.harvard.edu/MIST/interp_tracks.html}} \citep{dotter16, choi16} and a simulated stellar population from PARSEC\footnote{\url{http://stev.oapd.inaf.it/cmd}} \citep{bressan12}. 

We limit the data and the models to the same \teff-\logg range ($4900-5200$~K and $\logg < 2.5$, where both the RGB and AGB samples are expected to be most complete) and select a narrow range in observed metallicities ($\pm 0.1$~dex) around the model $\feh = -1.75$. In the data, $29\%$ of the stars lie above the dotted RGB/AGB separation line in Figure~\ref{fig:HR-carbon}. Within the selected parameter range in the models, stars spend $13\%$ of their time on the AGB on a MIST evolutionary track of a $0.8 M_{\odot}$, and $13\%$ of the stars in a 12 Gyr old PARSEC simulated population of $10^5$ M$_\odot$ with a \citet{kroupa01, kroupa02} IMF live on the AGB. Each of these numbers can change by a few percent depending on the chosen parameter limits. 

There is a difference in relative number of AGB stars between the models ($13\%$) and the data ($29\%$), but for a simple estimate such as this it is not unacceptable. First, our separation of the data into RGB or AGB star is very rough, and the introduced cuts in \teff and \logg for the comparison potentially bias the results. Furthermore, in reality, there is a range of ages and masses of stars in the inner Galaxy (and possibly helium abundances), complicating a model comparison. Additionally, evolutionary tracks and simulated stellar populations are not perfect. More detailed comparisons with for example observed globular clusters may be useful, but are beyond the scope of this work. Another possibility is that our photometric target selection favours the carbon-poor AGB stars over carbon-normal stars, due to weaker carbon features in the spectra, although this effect is likely too small. 

\subsection{Deriving evolutionary corrections}\label{sec:ccor}

The evolutionary carbon correction is stronger for stars with lower \logg and stars of lower \feh, which is visible in panel a) of Figure~\ref{fig:loggC}. It is less strong for very carbon-rich stars \citep{placco14}. These three parameters, \logg, \feh and \cfe, are the input for the derivation of the corrections. 

Including stellar parameter uncertainties in the derivation of the corrections is crucial, especially since the correction depends strongly on \logg and the PIGS uncertainties in \logg are non-negligible. For each star, we take 100 draws from normal distributions around each of the parameters with widths equal to their total uncertainties, and compute carbon corrections for each of them. The mean of these 100 carbon corrections is adopted as the final carbon-correction for each star. We checked what happened when only varying one of the stellar parameters while keeping the other two constant, from which we concluded that the carbon corrections are affected the most by including uncertainties for \logg, and the differences for \feh and \cfe are minor. 

The average \cfe as function of \logg for carbon-corrected RGB stars is shown in panel b) of Figure~\ref{fig:loggC}. The trend is largely flat, indicating that the carbon-corrections are properly determined. There appears to be a small dip around $\logg = 2.0$ (although it is consistent within the uncertainties), which potentially indicates contamination by AGB stars. 

The carbon corrections are only valid for RGB stars, they underestimate the depletion of carbon for AGB stars. We experimented with adopting tip-of-the-RGB \logg values for AGB stars (stars above the dashed line in Figure~\ref{fig:HR-carbon}) to derive the correction with those. However, we found that this typically over-estimates the carbon-correction: the AGB stars end up with a higher average \cfe than the RGB stars. Additionally, it is difficult to decide for each individual star whether it is an AGB or an RGB star. Finally, this analysis depends on the correctness of the isochrones regarding the location of the tip of the RGB. At low \logg, the carbon-correction is very sensitive to the exact \logg so small deviations have a large impact. It also depends on the assumption that no additional mixing happens between the tip of the RGB and early AGB, which is not known. Taking these considerations into account, we decide to not derive special corrections for the AGB stars but adopt the normal correction, considering it a lower limit. For VMP stars, the relative number of AGB stars is small. 

\begin{figure*}
\centering
\includegraphics[width=0.6\hsize,trim={0.0cm 0.0cm 0.0cm 0.0cm}]{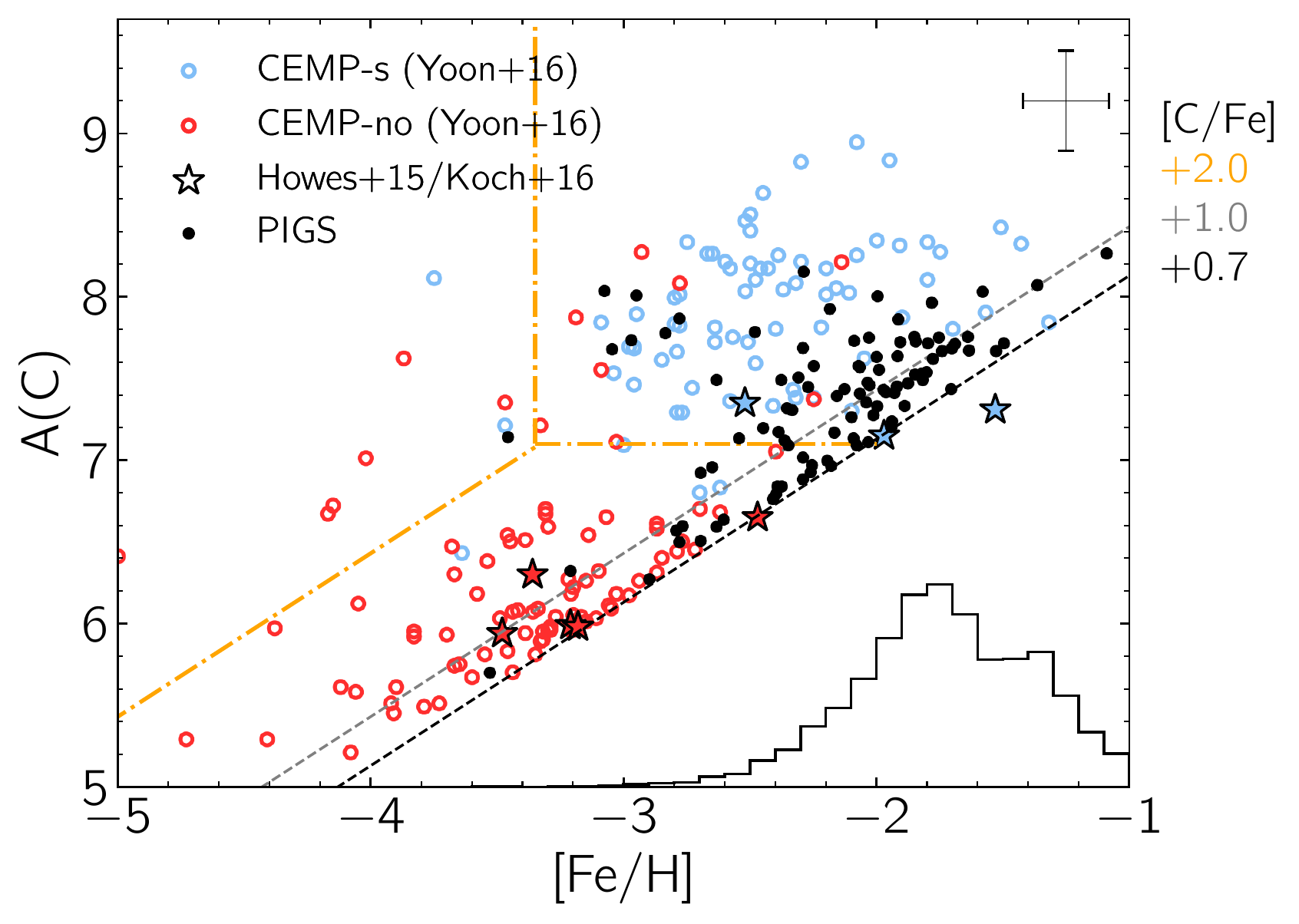} 
\caption{A(C)--\feh diagram for CEMP stars, with the new PIGS stars in black, a literature halo sample from \citet{yoon16} as red and blue circles (including only giant stars with $\teff < 5500$~K and $\logg < 3.5$), and the seven previously known bulge CEMP stars and one CH star from \citet{howes15} and \citet{koch16} with star symbols, coloured according to their CEMP class (see Section~\ref{sec:embla} for a discussion on the Howes et al. stars). Two dashed lines of constant \cfe are shown: $\cfe = +0.7$ in black is our CEMP criterion and $\cfe = +1.0$ (sometimes used in the literature) in grey. 
The dashed-dotted orange lines roughly separate the CEMP stars according to the \citet{yoon16} classification, with their Group I stars on the top right (mainly CEMP-s stars) and the Group III and II stars (mainly CEMP-no stars) on the top left and the bottom, respectively. These lines converge at A(C)~$= 7.1$, $\cfe = +2.0$ and $\feh = -3.35$.
The metallicity distribution function of the 8652 PIGS stars with $\feh < -1.0$ in the same \teff and \logg range as the CEMP stars is represented by the histogram. Typical uncertainties on the PIGS \feh and A(C) are shown in the top-right corner.}
    \label{fig:ac}
\end{figure*}

\section{CEMP stars in PIGS} \label{sec:cemp}

The main goal of this work is to investigate the CEMP stars in PIGS. In this section, we report the discovery of a large number of CEMP stars in the inner Galaxy. We also discuss the CEMP fraction as a function of the metallicity, compared to the halo. 

\subsection{New CEMP stars}\label{sec:newCEMP}

Throughout this work, the CEMP definition we adopt is $\cfe \geqslant +0.7$. Among a sample of 8652 cool metal-poor giant stars in PIGS with $\feh < -1.0$, $\teff < 5700$~K and $\logg < 3.5$ we identify 96 CEMP stars, 47 of which have $\cfe \geqslant +1.0$. For $\feh < -2.0$, these numbers are 62 and 36, respectively, out of 1836 stars. The \cfe was corrected for evolutionary effects as described in Section~\ref{sec:carbongiants}. All of the PIGS CEMP stars are new discoveries, and they increase the total sample of known CEMP stars in the inner Galaxy more than ten-fold. 

Previously, only two CEMP stars were reported in the bulge region: one CEMP-s star \citep{koch16} and one CEMP-no star \citep[in EMBLA,][]{howes15, howes16}. When applying \citet{placco14} evolutionary carbon corrections to the EMBLA sample and adopting the $\cfe \geqslant +0.7$ CEMP definition, we find that three more stars can be classified as CEMP, all with $\feh < -3.0$ and low [Ba/Fe] (hence these are CEMP-no stars). We consider two additional stars with $\feh = -1.97$ and $-2.47$ and $\cfe = +0.69$ as CEMP stars. The more metal-rich of these also has a high [Ba/Fe] and can therefore be classified as a CEMP-s star. This brings the total number of previously known inner Galaxy CEMP stars to seven. 

The PIGS stars passing the CEMP criterion are shown as black circles on the \feh--A(C)\footnote{$A$(C) $= \log{\epsilon (C)} = \log(N_C/N_H) + 12$} diagram in Figure \ref{fig:ac}. As detailed in the introduction of this paper, the different sub-populations of CEMP stars (CEMP-no and CEMP-s stars) show a typical separation in A(C) space. The previously known CEMP stars in the inner Galaxy (and a slightly more metal-rich CH~star from \citealt{koch16}) are represented by coloured star symbols in Figure~\ref{fig:ac}, together with a large sample of giant halo CEMP stars compiled by \citet{yoon16}. Most of the CEMP stars in PIGS are found around $\feh = -2.0$, largely because that is where the metallicity distribution function of the PIGS follow-up peaks (as shown by the histogram in the bottom right corner). 

The CEMP stars with $\feh < -2.0$ are spread relatively evenly across the giant branch, although they tend to have slightly lower \logg compared to the carbon-normal stars. The CEMP stars with $-2.0 < \feh < -1.5$, however, are mainly (23/32) located at relatively high temperatures and low gravities, in the early AGB regime. This could be the result of selection effects against carbon-rich stars in the PIGS photometric selection which are stronger for cooler and more metal-rich stars, although it is not clear that such effects would be strong enough to create such a difference. 

In the absence of s-process abundance information\footnote{There are two strong lines of Sr and one of Ba in the blue arm spectra, but due to the combination of low resolution, sometimes low S/N and very low metallicity, we cannot measure them for any of the stars with $\feh < -2.0$.} or any radial velocity monitoring for our stars, a different strategy is needed to classify the newly discovered CEMP stars. A general classification can be made based on the A(C) and \feh of the stars \citep{spite13, bonifacio15, yoon16}. 
The latter authors suggest three different groups of CEMP stars: the Group I stars largely contain CEMP-s stars and are found at high A(C) ($> 7.1$) and higher metallicity, and the Group II and III stars contain largely CEMP-no stars and are located at lower A(C) and/or lower metallicity. A rough separation of these groups has been indicated with dashed-dotted orange lines in Figure~\ref{fig:ac}. The authors showed that the contamination of CEMP-no stars in Group I and CEMP-s stars in Groups II and III is small, although there are some ``anomalous'' stars which they discuss in detail. They also discuss possible different origins for the Group II and III CEMP-no stars.
Others have used a higher A(C) value for the CEMP-s/no separation \citep{bonifacio15,lee19}, but those samples contain largely main-sequence turn-off stars whereas the \citet{yoon16} sample has a larger number of giants. Since PIGS contains solely giants, we adopt the \citet{yoon16} separation. We find that 72 stars lie in the CEMP-s region, and 24 in the CEMP-no part of the diagram. It is noteworthy that most of the CEMP-s candidates lie relatively close to the $\cfe = +0.7$ line, whereas the literature sample does not and has much higher \cfe values. Most of the CEMP-no candidates in PIGS also lie close to the $\cfe = +0.7$ line, but here it is more consistent with the distribution of halo CEMP-no stars. The star around $\feh = -3.5$ and A(C)~$=7.1$ is the only candidate Group III CEMP-no star. 

\begin{figure}
\centering
\includegraphics[width=1.0\hsize,trim={0.0cm 0.0cm 0.0cm 0.0cm}]{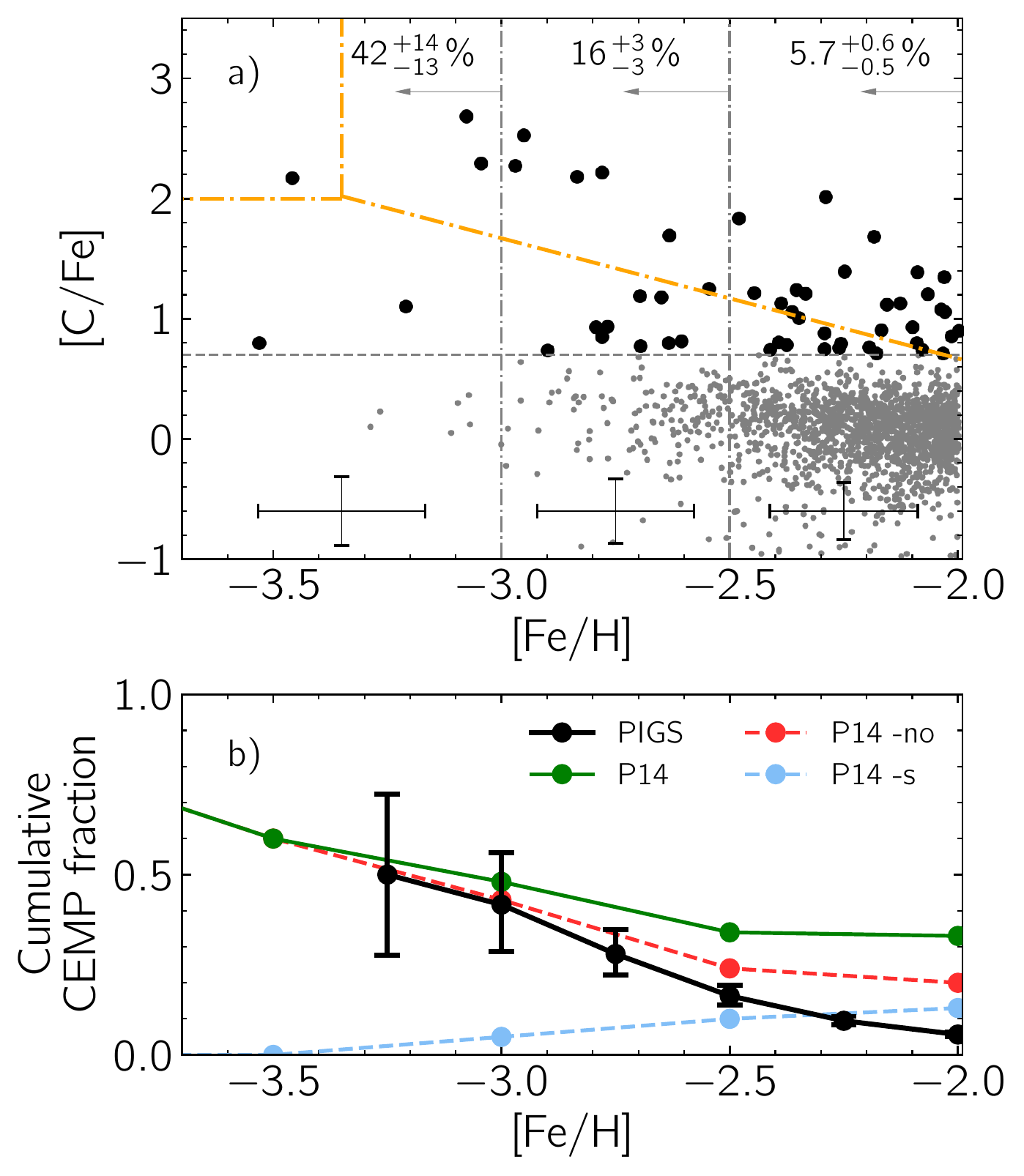} 
\caption{Top: \feh versus \cfe for PIGS stars with $4600$~K~$< \teff < 5500$~K. The horizontal dashed line indicates the CEMP criterion at $\cfe = +0.7$, the CEMP stars (which are above this line) are shown with larger black symbols. The orange dashed lines are the same as in Figure~\ref{fig:ac}. The PIGS cumulative CEMP fraction below different metallicities (vertical dot-dashed lines) is indicated in black. Typical uncertainties on \feh and \cfe are shown on the bottom of each metallicity range. Bottom: PIGS cumulative CEMP fraction as function of \feh for the same sample. The halo CEMP fraction from \citet{placco14} is shown in green, and separated into the fraction of CEMP-no (red dashed) and CEMP-s (blue dashed) stars. }
    \label{fig:fraction}
\end{figure}

\subsection{The CEMP fraction} \label{sec:CEMPfraction}

In the halo, the CEMP fraction is $\sim30 \%$ for stars with $\feh < -2.0$ and rises quickly for lower metallicity stars \citep[e.g.,][and references therein]{placco14}. For a quantitative comparison between the distribution of CEMP stars in PIGS and the Galactic halo, we derive the cumulative CEMP fraction as a function of metallicity. We select VMP stars in PIGS with $4600$~K~$< \teff < 5500$~K, a regime where the carbon abundances are expected to be the most reliable. Stars with $\epsilon_{\cfe} > 0.5$~dex are discarded. We draw 1000 samples from the distributions of \feh, \cfe and their uncertainties, and adopt the median of the 1000 CEMP fractions and uncertainties as our final values\footnote{The uncertainties on the fractions are computed using \citet{wilson} score confidence intervals of $1\sigma$, as in e.g. \citet{yoon18}.}.

The \cfe as a function of \feh for the selected sample is presented in panel a) of Figure~\ref{fig:fraction}, with the PIGS cumulative percentage of CEMP stars below different \feh values indicated in black. For $\feh < -3.0, -2.5, -2.0$, we find CEMP fractions of $42^{\,+14\,}_{\,-13} \%$, $16^{\,+3\,}_{\,-3} \%$, $5.7^{\,+0.6\,}_{\,-0.5} \%$, respectively. Panel b) presents a comparison of the cumulative CEMP fraction in PIGS with the halo fraction from \citet{placco14} (including the halo fraction separated for CEMP-s and CEMP-no stars). The PIGS fraction appears compatible with the overall literature fraction for $\feh < -3.0$. For $\feh < -2.0$, however, the CEMP fraction in PIGS is strongly discrepant with the halo fraction which is about six times larger. Table~\ref{table:cfrac} presents the cumulative CEMP fraction in PIGS for different carbon-enhanced definitions ($\cfe \geqslant +0.5, +0.7, +1.0$), for reference.

The orange line in panel a) of Figure~\ref{fig:fraction} can serve as a rough division between CEMP-s and CEMP-no stars. The CEMP stars with $-2.5 < \feh < -2.0$ mostly lie in the CEMP-s regime, but at lower metallicities the sample appears approximately equally split between CEMP-s and CEMP-no. However, the sample sizes are small and higher resolution spectroscopic follow-up is necessary to properly assign the stars to either category -- we therefore refrain from presenting separate CEMP-no and CEMP-s fractions here.

\renewcommand{\arraystretch}{1.5}
\begin{table}
\centering
\caption{\label{table:cfrac}PIGS cumulative CEMP fraction for different \cfe limits}
\begin{tabular}{lccc}
 \hline
 \feh         & $\cfe \geqslant +0.5$                           & $\cfe \geqslant +0.7$                         & $\cfe \geqslant +1.0$ \\
 \hline
$-3.25$   & $50.0^{\,+22.4\,}_{\,-22.4} \%$     & $50.0^{\,+22.4\,}_{\,-22.4} \%$    & $25.0^{\,+25.0\,}_{\,-15.0} \%$ \\
$-3.0$     & $50.0^{\,+13.9\,}_{\,-13.9} \%$     & $41.7^{\,+14.3\,}_{\,-13.0} \%$    & $33.3^{\,+14.4\,}_{\,-11.9} \%$ \\
$-2.75$   & $40.0^{\,+7.1\,}_{\,-6.7} \%$         & $28.0^{\,+6.7\,}_{\,-5.9} \%$        & $20.0^{\,+6.2\,}_{\,-5.0} \%$ \\
$-2.5$     & $28.1^{\,+3.6\,}_{\,-3.3} \%$         & $16.4^{\,+3.0\,}_{\,-2.6} \%$        & $8.8^{\,+2.4\,}_{\,-1.9} \%$ \\
$-2.25$   & $19.1^{\,+1.6\,}_{\,-1.5} \%$         & $9.5^{\,+1.2\,}_{\,-1.1} \%$          & $4.1^{\,+0.9\,}_{\,-0.7} \%$ \\
$-2.0$     & $13.3^{\,+0.8\,}_{\,-0.8} \%$         & $5.7^{\,+0.6\,}_{\,-0.5} \%$          & $2.1^{\,+0.4\,}_{\,-0.3} \%$ \\
 \hline
\end{tabular}
\end{table}

\subsection{Face-value distribution of PIGS CEMP stars}

PIGS has increased the number of known CEMP stars in the inner Galaxy by more than a factor of ten. However, the face-value distribution of candidate CEMP stars in PIGS on the \feh vs. A(C) diagram is discrepant compared to the halo. There is a lack of very carbon-rich stars with A(C) $> 8.0$, especially for \feh around $-2.5$ and higher -- in the regime of the CEMP-s stars.
The CEMP fraction in PIGS is lower than the fraction known from the Galactic halo, except for $\feh < -3.0$. If the PIGS selection is unbiased and the inner Galaxy CEMP fraction were the same as in the halo, we would have expected to find around 600 CEMP stars with $\feh < -2.0$ in PIGS ($33\%$ of 1900 stars), whereas we found only 62. The difference is less striking for $\feh < -2.5$, where we would have expected to find $\sim60$ CEMP stars instead of the 20 in PIGS. 

How many of the CEMP stars in PIGS could be ``accidental'' -- simply the outliers of a Gaussian \cfe distribution? We fit a Gaussian to the \cfe distribution in Figure~\ref{fig:fraction}, excluding $3\sigma$ outliers and stars above the RGB/AGB separation line in Figure~\ref{fig:HR-carbon}. For the higher metallicity stars ($ < -2.5 < \feh < -2.0$), the mean \cfe is 0.07 with a sigma of 0.20~dex. From such a distribution, one would expect only 0.1\% of the stars to have $\cfe \geqslant + 0.7$. For the number of stars in PIGS in this metallicity range used in the CEMP fraction (1545), this corresponds to 1.6 CEMP stars, compared to the 33 in PIGS. Between $-3.0 < \feh < -2.5$ we find a \cfe mean of 0.17 with a sigma of 0.25~dex. From such a Gaussian distribution, one would expect 2.6 CEMP stars out of 158 in this metallicity range, compared to 15 in PIGS. We therefore conclude that most CEMP stars in PIGS are real outliers of the \cfe distribution. 

We want to stress here that we caution against interpreting the PIGS CEMP distribution and fractions at face value. In the next section we will discuss several uncertainties and systematic biases which will have to be taken into account before a solid conclusion on the CEMP fraction of the bulge (and its comparison to the halo) can be considered definitive.

\section{Discussion}\label{sec:disc}

\subsection{Carbon bias and selection function}\label{sec:selfunc}

The selection of metal-poor stars in PIGS could be biased against CEMP stars, since carbon-rich stars have large molecular features in their spectra which affect the flux in the \CaHK narrow-band and in the employed broad-band filters. The effect of carbon on the Pristine \CaHK photometry and the PIGS colour-colour selection (example in Figure~\ref{fig:ccd}) in particular, are investigated in detail in Arentsen et al. (in prep.). In that study, synthetic photometry is derived from synthetic spectra with different carbon abundances to test the behaviour of carbon-rich stars. However, a direct comparison between the synthetic photometry and our PIGS selection is complicated. There are important caveats regarding the synthetic photometry due the assumption of local thermodynamic equilibrium (LTE) in one dimension (1D) for the stellar spectra, and simplified assumptions about the chemical compositions of the stars. Additionally, it is difficult to reconstruct the PIGS selection function. This means that even if it was perfectly known how carbon-rich stars behaved in the colour-colour space, there is still some uncertainty to which stars would and would not have been selected for follow-up. Below, we will briefly discuss the results from the synthetic photometry analysis and describe several challenges regarding the reconstruction of the PIGS selection function.

\subsubsection{Carbon-rich photometry}

Since there are more absorption features of carbon in the blue part of the spectrum, carbon-rich stars appear redder than carbon-normal stars (this effect has been known for a long time, e.g. \citealt{bondneff69}). In our synthetic photometry analysis, we indeed find that a CEMP star moves to the right in our colour-colour diagram, and could even move out of our selection box in that direction. The differences are larger for stars with higher carbon abundances and/or cooler temperatures (especially strong effects are seen for $\teff < 5000$~K and $\cfe > +1.5$). We also find that CEMP-s stars are affected more than CEMP-no stars, due to different relative CNO abundances. 

In the y-axis of the colour-colour selection diagram, the behaviour of carbon-rich stars is more complicated (and different for CEMP-no and CEMP-s stars) due to the combination of three or four different filters. However, it appears that we have been ``lucky'' in selecting the colour combination in such a way that carbon-rich stars typically do not move down very much in our PIGS selection diagram. The synthetic stars are not leaking out of our selection box in that direction, but this can depend on the assumptions made. 

Finally, since we applied a fixed magnitude cut and carbon-rich stars are fainter in the blue, there can be a slight selection bias against the faintest CEMP stars. This is expected to be worse for the PS1 selection where the magnitude cut is based on the bluer $g$ magnitude (4000 \AA~$-$~5500 \AA) compared to the \Gaia selection where the $G$ magnitude was used (which covers a much wider wavelength range, from 3400 \AA~$-$~10\,000 \AA). 

\subsubsection{PIGS selection function}

There are multiple challenges in reconstructing the selection function of PIGS follow-up targets. First, we wish to emphasise again that the high extinction towards the inner Galaxy makes studying this region (and determining a selection function) extremely challenging. The population selected for follow-up in PIGS will vary from field to field depending on the extinction, with (for example) varying contamination by more metal-rich stars. The extinction correction is especially difficult for the wide \Gaia filters, but the PS1 correction is also far from perfect. The extinction map used \citep[][or earlier versions]{green19} has variable quality throughout the footprint, has limited resolution, is affected by three-dimensional effects that are not taken into account in our analysis, and adopts a single extinction law (whereas it may in fact be variable). 

Other factors that determine which $\sim 360$ stars per 2dF field are selected for follow-up are, for example, the underlying metallicity distribution and the stellar density, which vary from field to field and depend strongly on the Galactic latitude. Additionally, not all the best targets are actually assigned a fibre in the fibre allocation process, since fibres can only be placed a certain distance from each other. The allocation also depended on the availability of \CaHK photometry in a field (which was not always homogeneous) and the inhomogeneity of the extinction. 

Furthermore, the reconstruction of our selection function is hampered by various changes throughout the survey. The goal was to keep improving our yields of the most metal-poor stars in the inner Galaxy, which is why these changes were made, but they result in an inhomogeneous survey selection. For example, we switched from \Gaia to PS1 due to extinction challenges. The selection with the two surveys used different magnitude ranges, which roughly correspond to each other (but not perfectly). The discrepancy is higher for higher extinctions, since $g$ is much more affected than $G$. Other changes were made throughout the survey in the colour limits of the selection, in cuts on the parallax, in the way the extinction was treated, in the zero-point calibration of the \CaHK photometry, and in the removal of variable stars. 

\subsubsection{Halo Pristine samples}

Other \Pristine-selected samples also have low CEMP fractions among giant stars (\citealt{aguado19} corrected for \logg effects, \citealt{caffau20}, Lucchesi et al. in prep.). However, these samples were selected much more strictly than the PIGS stars: stars were observed one by one based on their low photometric metallicities and not with a box selection as in PIGS. The selection effects are expected to be larger in the halo samples (this will be presented in more detail in Arentsen et al. in prep.). 

\subsubsection{Summary}

There are several complications in reconstructing the PIGS selection function. Combined with the uncertainties on the modelling of synthetic spectra for CEMP stars (and hence uncertainties on synthetic photometry), we conclude that it is not possible to make solid predictions about which stars should or should not have ended up in our final PIGS follow-up sample. It is likely that some of the coolest, most carbon-enhanced stars were excluded from our selection, but it is not at all clear that this can fully explain the low CEMP fraction in PIGS. The generous box-selection of PIGS still contains many of the carbon-rich stars.

\subsection{Comparison with the literature}

\subsubsection{Halo samples}\label{sec:disc-plac}

Throughout this work, we have used the \citet{placco14} CEMP fraction as a reference. It is based on a compilation of $\sim 500$ VMP stars studied with high-resolution spectroscopy, which was expected to be unbiased with respect to carbon. The sample includes stars of various evolutionary phases, whereas PIGS contains only giant stars. To test whether there is a difference in the fraction between giants and turn-off/main sequence stars, we re-derive the CEMP fraction from the \citet{placco14} sample for giants only (2/3rd of the sample). For stars with $\teff < 5500$~K and $\logg < 4.0$, the fractions for $\feh < -2.0, -2.5, -3.0$ go down by $5\%, 5\%, 7\%$, respectively, to $28\%, 29\%, 41\%$. This does not solve the tension between PIGS and the \citet{placco14} fraction.

What about lower resolution spectroscopy studies in the halo? There have been a number of papers in recent years studying large  samples of halo metal-poor stars at low/intermediate resolution. These samples consist of largely giant stars, and the authors typically report overall CEMP fractions in agreement with the high-resolution \citet{placco14} fraction \citep[e.g.][]{placco18,placco19,yoon18,limberg21}. However, they sometimes do have a lack of the most carbon-rich CEMP stars, and they have varying overall distributions of CEMP stars on the \feh -- A(C) diagram. It is beyond the scope of this paper to investigate these differences in detail, but they may be important. 

\subsubsection{High-resolution inner Galaxy EMBLA sample}\label{sec:embla}

The metal-poor inner Galaxy SkyMapper/EMBLA survey reported only one extremely metal-poor CEMP-no star with $\cfe > +1.0$ and zero CEMP-s stars among their 33 stars with measured carbon abundances \citep{howes15,howes16}. We identified five additional CEMP stars in their sample (see Section~\ref{sec:newCEMP}). 

Including the new CEMP stars, we find a CEMP fraction of 44\% (4/9) for $\feh < -3.0$, consistent with the halo CEMP fraction of 48\% \citep{placco14} and the PIGS fraction of 42\%. For $\feh < -2.5$, the CEMP fraction is 18\% (4/22), consistent with the PIGS fraction of 16\% and the halo CEMP-no fraction of 24\%, but lower than the overall CEMP halo fraction of 34\%. For $\feh < -1.95$, including the two almost-CEMP stars, the CEMP fraction is 20\% (6/30), lower than the halo fraction of 33\% but higher than the PIGS fraction of 6\%. Only considering CEMP-no stars, the fraction is 17\% (5/29), consistent with the CEMP-no halo fraction of 20\%. 

The EMBLA sample is on average relatively cool with almost all stars having $\teff < 5000$~K, which is where photometric selection effects are stronger. However, due to their similar ``box''-selection strategy as in PIGS, stars with moderate carbon-enhancement likely would not have been missed. Their lack of very carbon-rich CEMP stars (mainly CEMP-s) is potentially due to selection effects -- the highest \cfe in EMBLA is +1.23, all other CEMP stars have $\cfe < +1.0$. The CEMP-no fraction in EMBLA appears to be consistent with the CEMP-no fraction in the literature, but the CEMP-no stars in the literature have on average higher carbon abundances compared to those in EMBLA. 

\subsection{Abundance variations and 3D/non-LTE}

The exact distribution of A(C) in the PIGS CEMP stars can be affected by assumptions in the spectroscopic analysis. In the models used in the FERRE analysis, oxygen follows the alpha abundances and therefore [O/Fe] $= +0.4$, while nitrogen follows the solar abundance pattern and therefore [N/Fe] $= 0.0$. In observed CEMP stars, there is a range of relative CNO abundances. For most stars this does not have a significant impact on the spectra as the strongest carbon feature is the CH G-band, but for cool and very carbon-rich (e.g. $\cfe > +2.0$) stars other features such as CN and C$_2$ start to become larger as well, and these depend on the CNO assumptions. Inspecting the fits of high A(C) stars we noticed that the C$_2$ features were often weaker in the models compared to the data, although the CH G-band was well fitted. This potentially affects the FERRE \cfe derivation for such stars. 

The synthetic grid we used for the analysis follows the simple 1D/LTE assumptions, which are usually also adopted for the analysis of CEMP stars in the halo. Recently, \citet{norrisyong19} have cast doubts on the reported CEMP(-no) fractions in the literature. They found that once estimates for 3D and non-LTE corrections on iron and the CH G-band are taken into account (which are stronger for more metal-poor stars), the CEMP star carbon abundances decrease by up to $\sim0.5$~dex, reducing the number of CEMP-no stars by $\sim 70\%$. Similar results were found by \citet{amarsi19}, who reported lower carbon abundances in a careful 3D/non-LTE analysis of atomic carbon for a number of metal-poor stars. Our estimated CEMP fraction can also be affected by these effects. However, since the CEMP fractions in the halo and in PIGS are both based on similar assumptions, they can likely be compared without problem. One difference is that PIGS consists exclusively of giants, whereas literature samples are a mixture of dwarfs, turn-off stars and giants. These can be expected to be affected differently by 3D/non-LTE effects. However, even when limiting the high-resolution halo sample to only giants, there is a large discrepancy in the CEMP fractions (see the previous section).

\subsection{Interpretation of the PIGS CEMP population}

How \emph{can} we interpret the population of CEMP stars in PIGS? There are strong differences between the CEMP fractions in PIGS and those known from the literature in the Galactic halo, which potentially reflect truths about the underlying stellar population and/or be related to selection effects. We discuss some possible interpretations for the CEMP-s and the CEMP-no stars.

\subsubsection{CEMP-s stars}

CEMP-s stars are expected to have interacted with a binary companion in the past, where a former AGB star has transferred carbon-rich material to the star we today see as carbon-enhanced. In the halo, roughly $10-13\%$ of the stars with $\feh < -2.0$ are CEMP-s stars \citep{placco14,yoon18}. In PIGS, the distinction between CEMP-s and CEMP-no is not trivial because we have no s-process abundance measurements. A rough separation can, however, be achieved by selecting stars with high A(C) to be CEMP-s, and stars with low A(C) as CEMP-no, as in Section~\ref{sec:newCEMP}. We have seen that there is a strong discrepancy between the CEMP population in PIGS and the halo in the A(C) $>8.0$ regime. The PIGS sample has almost none of these stars, whereas they are abundant in the halo. The \emph{fraction} of CEMP stars for $\feh < -2.0$ and $< -2.5$ is also discrepant with that in the halo. For $\feh < -2.0$, the overall CEMP fraction in PIGS is only $6\% \pm 0.5\%$ (see Figure~\ref{fig:fraction}). Half of the CEMP stars in this range have A(C) $> 7.1$ and hence are CEMP-s by this definition. 

The discrepancy with the halo CEMP-s fraction and the lack of stars with very high A(C) is potentially partly due to photometric selection effects, which are strongest at higher metallicities and higher carbon abundances. However, it is not certain that the selection effects are strong enough to fully account for such a low fraction. Based on the bias analysis in Arentsen et al. (in prep.), only a small percentage of CEMP-s stars is expected to move significantly in the colour-colour diagram in such a way as to fall outside of our photometric selection box. We might also have expected a higher CEMP fraction among warmer stars in our sample due to selection effects, which we do not observe. 

Assuming that the photometric selection cannot fully account for the missing CEMP-s stars, there appears to be a real lack of CEMP-s stars in the inner Galaxy. 
According to the population synthesis modelling of \citet{abate15b}, there are several parameters that could change the fraction of CEMP-s stars by up to a few per cent; for example variations in the initial mass function, variations in the initial mass ratio and the period distribution of binary stars and the age of the population. Naturally a different binary fraction of the stars would also result in a difference in the number of CEMP-s stars. Not much is known about the number and the properties of binary stars in the inner Galaxy, let alone at the lowest metallicities.

The binary fraction of stars is expected to be lower in very high-density environments (mainly because of dynamical interactions) -- this has been observed in globular clusters, which appear to have lower binary fractions compared to the field \citep[e.g.][]{milone12,lucatello15}. \citet{dorazi10} studied the occurrence of barium-rich stars (Ba/CH/CEMP stars) in globular clusters of various metallicities, and found a much lower fraction of barium-rich stars in clusters compared to the field. Their sample contains five clusters with $\feh \leq -2.0$, with 50--100 stars per cluster, and they find only one barium-rich star among all these. There appears to be a low CEMP-s fraction in globular clusters indeed.

The bulge population nowadays is not as dense as a globular cluster, but recent observations suggest that the contribution of dissolved globular clusters to the halo is significantly larger in the inner few kpc of the Milky Way than further out \citep{schiavon17,horta21}. The latter authors estimate that the fraction of the mass originating from globular clusters is $\sim 27\%$ within 1.5~kpc, six times higher than at 10~kpc. Simulations also show that the fraction of stars in the inner Galaxy originating from globular clusters may be relatively large \citep{hughes20}. If many of the $\feh < -2.0$ inner Galaxy stars originated in globular clusters, like their more metal-rich counterparts at $-2.0 < \feh < -1.0$ in \citet{horta21}, their binary fraction can be lower than that of outer halo stars. This results in a lower CEMP-s fraction. 
 
\citet{yoon18} and \citet{lee19} found that the \emph{local} inner halo (close to the Sun) has relatively more CEMP-s stars than CEMP-no stars compared to the outer halo, but this possibly only represents the fact that the local inner halo is more metal-rich than the outer halo. They do not compare the fraction of CEMP-s stars between the inner and outer halo in a smaller metallicity interval (e.g. $-2.5 < \feh < -2.0$).

\subsubsection{CEMP-no stars}

There appears to be agreement between the CEMP fraction in PIGS and the overall literature fraction for $\feh < -3.0$, although we note that this is based on a relatively small sample of $\sim 15$ stars with unknown CEMP classes. Most literature CEMP stars in this metallicity range are CEMP-no stars: \citet{placco14} report a CEMP-no fraction of 43\% and a CEMP-s fraction of only 5\%. In PIGS, three of the five CEMP stars in this range are most likely to be CEMP-no stars (when including the star very close to the A(C) division), but the other two overlap more with the CEMP-s stars in the \feh -- A(C) diagram. Excluding the stars with A(C) $> 7.5$, the CEMP-no fraction for EMP stars is $25^{\,+14\,}_{\,-10} \%$, inconsistent with the literature CEMP-no fraction. 

For $\feh < -2.5$, the fraction of CEMP-no stars in the Galactic halo is $24\%$ \citep{placco14}, and the total fraction of CEMP stars in PIGS is $21\%$. However, roughly half of the PIGS CEMP stars in this range have A(C) $> 7.1$ and likely belong to the CEMP-s category. The fraction of CEMP-no stars in this metallicity range therefore appears to be lower in PIGS than expected from the literature for the halo. For $\feh < -2.0$ this is most certainly the case, where $20\%$ of the stars are expected to be CEMP-no but the overall fraction in PIGS is only $6\%$ (and some of these can be CEMP-s stars). 

Stars with $\feh < -3.0$ are not expected to be significantly affected by carbon-biases in the photometry, except for the coolest ($\teff < 4750$~K) and most carbon-rich ($\cfe > +2.0$) stars (Arentsen et al in prep.). Stars of these temperatures are only a minority in the PIGS sample, and stars with such high carbon abundances are a minority among the literature CEMP-no stars. It can therefore be expected that the CEMP-no fraction in this metallicity range is hardly affected by a photometric bias. For stars with $\feh < -2.0$ and $< -2.5$, slightly warmer and less carbon-rich stars are also affected, but still most of the CEMP-no stars are expected to stay within our selection.

Our observations suggest that the CEMP-no fraction in the inner Galaxy is inconsistent with the halo, with the discrepancy becoming larger at higher metallicities. The EMBLA inner Galaxy CEMP-no fraction may appear consistent with the halo, but their CEMP-no stars all have relatively low \cfe. Assuming possible selection biases are not the (sole) cause of the PIGS and/or EMBLA discrepancies, we search for a process that reduces the inner Galaxy CEMP-no fraction, primarily at higher metallicities. The stars that ended up in the bulge region possibly formed in larger building blocks (which are able to retain more of their gas) compared to stars ending up in the diffuse halo (likely born in smaller building blocks, which quickly lose most of their gas). Chemical evolution proceeded more quickly in dense regions, which can for example be seen in the chemical abundances of bulge stars \citep[for a review, see e.g.][]{barbuy18}. The C-enhanced signature from the First Stars is potentially still visible in the EMP stars because those formed very early on, but wiped out at higher metallicities due to the rapid chemical enrichment and mixing of the ISM with the ejecta from many Population II supernovae. This reduces the CEMP-no fraction at higher metallicities in the inner Galaxy. Additionally, to our best knowledge there are currently no CEMP-no stars known in globular clusters. If there has been a larger contribution of globular cluster stars to the inner Galaxy compared to the halo, as discussed in the previous section, this also reduces the CEMP-no fraction. 

\subsection{Future work}\label{sec:future}

This work represents a significant step forward in the study of the population of CEMP stars in the inner Galaxy. We conclude that deriving unbiased CEMP fractions based on photometrically selected samples is an extremely challenging task. There are several avenues to investigate the inner Galaxy CEMP population further. 

First, it will be crucial to do high-resolution spectroscopic follow-up of our CEMP candidates to get their detailed abundance patterns which will reveal information about their origin. Stars that became carbon-enhanced due to mass-transfer from a binary companion are expected to have high s-process abundances (e.g. strontium and barium). CEMP-no stars have typical abundance patterns which can be related to the properties of the First Stars and their supernovae that enriched the medium out of which they formed. The high A(C) stars around $\feh = -3.0$ are of particular interest, as they lie in a region of parameter space that holds both CEMP-no and CEMP-s stars. Curiously, \citet{arentsen19_cemp} have shown that binarity among CEMP-no stars in this A(C) regime is higher than among the CEMP-no stars with lower A(C) (although this still has to be confirmed with larger samples). They hypothesise that some of these stars were born with less carbon initially, but have obtained additional carbon as their binary companion transferred it to them. These CEMP-no stars do not show the typical mass-transfer s-process abundance enhancements, but it is questionable whether AGB stars at such low metallicities do produce s-process elements in the same way stars at higher metallicities do (see \citealt{arentsen19_cemp}, and references therein). Therefore not only should the abundance patterns of our stars be determined, monitoring the radial velocity adds valuable information as well. 

Secondly, we have not taken into account the distances and/or orbits of the stars in our sample. We have implicitly assumed that all of them are inner Galaxy stars, loosely defined as stars in the inner $4-5$ kpc of the Milky Way. However, some stars are potentially simply passing through the inner regions on highly elliptical orbits with large apocentres and are thus not truly inner Galaxy stars. It is still unclear how many of the metal-poor stars currently in the inner Galaxy are halo interlopers, with estimates ranging from $25 - 75\%$ \citep{howes15, kunder20, lucey21}. It is desirable to have orbital information of all the stars in PIGS, to be able to select a true inner Galaxy sample. We will investigate the orbital properties of PIGS stars in a future work. Unfortunately, the inner Galaxy is too far away to have good parallax information in the latest \Gaia data release \citep[EDR3,][]{gaiaedr3}, which makes distance determinations challenging. Better distances can be estimated if photometric information is included, but this is challenging towards the bulge due to the high, inhomogeneous extinction. Additionally, this will be especially difficult for CEMP stars since their colours are affected by carbon. 

Additional work on the radial variation in the CEMP fraction in the inner Galaxy and the rest of the halo and the relative occurrence of globular cluster stars is necessary to shed more light on the CEMP star population and what it can teach us about the build-up of the Milky Way. We do not study the radial variation in this paper, due to complications with the selection function of PIGS and the photometric bias for carbon-rich stars combined with the high extinction and unknown distances.

Finally, the details of the photometric selection of metal-poor candidates should be further investigated. The synthetic analysis of Arentsen et al. (in prep.) to test biases against carbon-rich stars is useful, but has its limits. There are several observational avenues that can shed light on the situation. The main \Pristine survey will perform a large follow-up campaign using WEAVE (the William Herschel Telescope Enhanced Area Velocity Explorer, a multi-object spectroscopic facility in the Northern hemisphere, \citealt{weave}), that will result in an incredibly rich spectroscopic sample of very metal-poor stars, including measured carbon and s-process abundances. WEAVE does not cover the inner Galaxy, but the Southern hemisphere 4-metre Multi-Object Spectroscopic Telescope (4MOST, \citealt{4most}) will. The 4MIDABLE-LR consortium survey will target metal-poor stars in the inner Galaxy \citep{chiappini19}, partly pre-selected from PIGS. This will significantly enlarge the number of very metal-poor inner Galaxy stars with spectroscopic follow-up, including estimates for carbon and s-process abundances useful for the study of CEMP stars. The WEAVE and 4MOST samples may still be biased with respect to carbon, but the homogeneous selection, sample size and data quality should allow for a good quantification of the effect. A further promising avenue for producing less biased samples of CEMP stars may be the use of a set of photometric filters with the sensitivity to measure carbon, for example using a narrow-band filter covering the G-band. A first encouraging exploration of this approach has been done by \citet{whitten19, whitten21}, for the J-PLUS and S-PLUS surveys \citep{cenarro19, mendes19}. The dispersed BP/RP spectra in future \Gaia releases may also be very useful for such an investigation.

\section{Summary and conclusions}\label{sec:conc}

In this paper, we have studied the occurrence of CEMP stars in the inner Galaxy using the \Pristine Inner Galaxy Survey (PIGS). Our main results can be summarised as follows:

\begin{itemize}
	\item The final PIGS spectroscopic sample after three years of follow-up contains 6700 stars with $\feh < -1.5$, of which 1900 have $\feh < -2.0$. This is the largest sample of very metal-poor stars in the inner Galaxy to date, and is well-suited for the study of CEMP stars in the central regions of the Milky Way. 
	\item The carbon abundance of RGB stars in PIGS decreases with decreasing \logg, which is also seen in halo stars. The magnitude of the carbon depletion is consistent with previous models by \citet{placco14}. We used those models to derive carbon corrections for the entire PIGS sample. 
	\item We discovered 96 new CEMP stars in PIGS, which is more than a ten-fold increase in the number of known CEMP stars in the inner Galaxy, since only seven CEMP stars were previously found in the inner Galaxy \citep{howes15, howes16, koch16}. Of the PIGS CEMP stars, 72 lie in the CEMP-s range of the \feh --A(C) diagram, and there are 24 that are likely CEMP-no stars. 
	\item The fraction of CEMP stars with respect to carbon-normal stars is lower in PIGS than in Galactic halo literature samples, except for stars with $\feh < -3.0$. As a function of metallicity, the cumulative CEMP fractions in PIGS are $42^{\,+14\,}_{\,-13} \%$ ($\feh < -3.0$), $16^{\,+3\,}_{\,-3} \%$ ($\feh < -2.5$) and $5.7^{\,+0.6\,}_{\,-0.5}$ ($\feh < -2.0$). The CEMP stars that are lacking compared to halo samples are mainly the more metal-rich and more carbon-rich stars. 
	\item We concluded that there are many uncertainties in the selection function of PIGS and in estimates of photometric selection biases against carbon-rich stars, hindering a conclusive prediction regarding which CEMP stars would have ended up in our selection and which would not have. To our best understanding, however, the carbon bias appears to be insufficient to fully explain the low CEMP fraction in PIGS.
	\item We speculate that a low fraction of CEMP-no stars (especially at higher metallicities) is potentially related to very quick chemical evolution in the high-mass building blocks that made up the inner Galaxy. A low fraction of CEMP-s stars in the inner Galaxy may be related to a lower binary fraction than in the rest of the halo, possibly due to a larger contribution from former globular cluster stars. 
	\item In the future, more light will be shed on the properties of CEMP stars in the inner Galaxy through high-resolution follow-up of our new CEMP stars, orbital analysis of PIGS stars, further tests of biases in the \Pristine selection, and through large samples of metal-poor inner Galaxy stars observed with 4MOST. 
\end{itemize}

The work presented in this paper is only the start of a promising avenue studying the build-up of the old, metal-poor inner Galaxy using carbon-enhanced metal-poor stars. 

\section*{Acknowledgements}

	We thank the Australian Astronomical Observatory, which have made these observations possible. We acknowledge the traditional owners of the land on which the AAT stands, the Gamilaraay people, and pay our respects to elders past and present. Based on data obtained at Siding Spring Observatory (via programs S/2017B/01, A/2018A/01, OPTICON 2018B/029 and OPTICON 2019A/045, PI: A. Arentsen and A/2020A/11, PI: D. B. Zucker). 
	
	Based on observations obtained with MegaPrime/MegaCam, a joint project of CFHT and CEA/DAPNIA, at the Canada-France-Hawaii Telescope (CFHT) which is operated by the National Research Council (NRC) of Canada, the Institut National des Science de l'Univers of the Centre National de la Recherche Scientifique (CNRS) of France, and the University of Hawaii.

	AA and NFM gratefully acknowledge funding from the European Research Council (ERC) under the European Unions Horizon 2020 research and innovation programme (grant agreement No. 834148).
	NFM, VH and GK gratefully acknowledge support from the French National Research Agency (ANR) funded project ``Pristine'' (ANR-18-CE31-0017). 
	DA thanks the Leverhulme Trust for financial support. 
	DBZ acknowledges the support of the Australian Research Council through Discovery Project grant DP180101791.
	The work of VMP is supported by NOIRLab, which is managed by the Association of Universities for Research in Astronomy (AURA) under a cooperative agreement with the National Science Foundation. 
	JIGH acknowledges financial support from the Spanish Ministry of Science and Innovation (MICINN) project AYA2017-86389-P. 
	Horizon 2020: This project has received funding from the European Union's Horizon 2020 research and innovation programme under grant agreement No 730890. This material reflects only the authors views and the Commission is not liable for any use that may be made of the information contained therein.

	The authors thank the International Space Science Institute, Bern, Switzerland for providing financial support and meeting facilities to the international team ``Pristine''. 
	
	This work has made use of data from the European Space Agency (ESA) mission {\it Gaia} (\url{https://www.cosmos.esa.int/gaia}), processed by the {\it Gaia} Data Processing and Analysis Consortium (DPAC, \url{https://www.cosmos.esa.int/web/gaia/dpac/consortium}). Funding for the DPAC has been provided by national institutions, in particular the institutions participating in the {\it Gaia} Multilateral Agreement. 

	The Pan-STARRS1 Surveys (PS1) and the PS1 public science archive have been made possible through contributions by the Institute for Astronomy, the University of Hawaii, the Pan-STARRS Project Office, the Max-Planck Society and its participating institutes, the Max Planck Institute for Astronomy, Heidelberg and the Max Planck Institute for Extraterrestrial Physics, Garching, The Johns Hopkins University, Durham University, the University of Edinburgh, the Queen's University Belfast, the Harvard-Smithsonian Center for Astrophysics, the Las Cumbres Observatory Global Telescope Network Incorporated, the National Central University of Taiwan, the Space Telescope Science Institute, the National Aeronautics and Space Administration under Grant No. NNX08AR22G issued through the Planetary Science Division of the NASA Science Mission Directorate, the National Science Foundation Grant No. AST-1238877, the University of Maryland, Eotvos Lorand University (ELTE), the Los Alamos National Laboratory, and the Gordon and Betty Moore Foundation.
	
	This research made extensive use of the \textsc{matplotlib} \citep{hunter07}, \textsc{pandas} \citep{pandas}, \textsc{astropy} \citep{astropy13,astropy18} and \textsc{dustmaps} \citep{dustmaps} Python packages, and of \textsc{Topcat} \citep{topcat}. 


\section*{Data Availability}

The data underlying this article will be shared on reasonable request to the corresponding author.


\bibliographystyle{mnras}
\bibliography{mpbulgeIII.bib}   



\bsp	
\label{lastpage}
\end{document}